\documentclass[useAMS,usenatbib]{mn2e}

\usepackage[pdftex]{graphicx}
\usepackage{aas_macros}
\usepackage{amsfonts,amssymb,amsmath}
\usepackage{times,helvet,varioref,color,epsfig}
\usepackage[
    pdftex,
    a4paper,
    plainpages=true,
    pdfpagelabels,
    breaklinks=true,
    bookmarks=true,
    bookmarksopen=true,
    bookmarksopenlevel=1
]{hyperref}

\hypersetup{
  pdfauthor = {Pascal J. Elahi},
  pdfkeywords = {methods: $N$-body simulations -- methods: numerical -- galaxies: haloes -- galaxies: subhaloes -- dark matter},
  pdftitle = {Subhaloes in Scale-Free Cosmologies}
}

\newcommand{\Msun}{\ {\rm M_\odot}}
\newcommand{\neff}{n_{\text{eff}}}

\newcommand{\Eqref}[1]{Eq.~(\ref{#1})}
\newcommand{\Figref}[1]{Fig.~\ref{#1}}
\newcommand{\Secref}[1]{\S\ref{#1}}
\newcommand{\Tableref}[1]{Table \ref{#1}}

\defcitealias{nfw}{NFW}
\defcitealias{shethtormen1999}{ST}
\defcitealias{pressschechter}{PS}

\title{Subhaloes in Scale-Free Cosmologies}
\author[P.J.~Elahi, R.J.~Thacker, L.M.~Widrow, and E.~Scannapieco]
       {Pascal~J.~Elahi$^1$, Robert~J.~Thacker$^2$, Lawrence~M.~Widrow$^1$ and Evan~Scannapieco$^3$\\
        $^1$Department of Physics, Engineering Physics \& 
Astronomy, Queen's University, ,
            Kingston, Ontario, Canada\\
        $^2$Department of Astronomy \& Physics, Saint Mary's University,
            Halifax, Nova Scotia, Canada\\
        $^3$School of Earth \& Space Exploration, Arizona 
State University, PO Box 871404,
            Tempe, Arizona, USA}

\begin{document}
\date{Accepted, Received; in original form }
\pagerange{\pageref{firstpage}--\pageref{lastpage}} \pubyear{2008}
\maketitle

\begin{abstract}
We explore the dependence of the subhalo mass function on the spectral index $n$ of the linear matter power spectrum using scale-free Einstein-de Sitter simulations with $n=-1$ and $n=-2.5$.  We carefully consider finite volume effects that may call into question previous simulations of $n<-2$ power spectra. Subhaloes are found using a 6D friends-of-friends algorithm in all haloes originating from high-$\sigma$ peaks. For $n=-1$, we find that the cumulative subhalo mass function is independent of the parameters used in the subhalo finding algorithm and is consistent with the subhalo mass function found in $\Lambda$CDM simulations. In particular, the subhalo mass function is well fit by a power-law with an index of $\alpha=-0.9$, that is the mass function has roughly equal mass in subhaloes per logarithmic interval in subhalo mass. Conversely, for $n=-2.5$, the algorithm parameters affect the subhalo mass function since subhaloes are more triaxial with less well defined boundaries. We find that the index $\alpha$ is generally larger with $\alpha\gtrsim-0.75$. We infer that although the subhalo mass function appears to be independent of $n$ so long as $n\gtrsim-2$, it begins to flatten as $n\rightarrow-3$. Thus, the common practice of using $\alpha\approx-1.0$ may greatly overestimate the number of subhaloes at the smallest scales in the CDM hierarchy.
\end{abstract}
\label{firstpage}

\section{Introduction}
\label{sec:intro}
In the current Cold Dark Matter (CDM) paradigm, structure forms hierarchically; small-scale density fluctuations collapse to form an early generation of haloes, which are the progenitors of larger-scale systems. While some progenitors survive as substructure, others are tidally disrupted and become the smooth component of the new system. The distribution of substructure within galaxy-size haloes is of great interest as galaxies and satellites, acting as baryonic tracers of the underlying mass distribution, can provide observational checks of the current CDM paradigm. The subhalo distribution at much smaller scales, likely devoid of baryonic tracers, is of practical interest for DM detection experiments.

\par
In this paper, we examine the properties of subhaloes in a pair of scale-free cosmologies meant to mimic two different scales in the Universe. Objects in a scale-free simulation can be related to objects at different scales in the CDM hierarchy via the scale dependence of effective spectral index $\neff\equiv d\ln P(k)/d\ln k$ of the $\Lambda$CDM concordance model. We use scale-free simulations for their simplicity as there is only one physical scale in the simulation, the nonlinear scale. This feature allows us to examine whether halo substructure remembers initial conditions, namely the index of the initial power spectrum.

\par
Dark matter haloes in $\Lambda$CDM simulations exhibit properties of self-similar and fractal systems. For example galactic haloes appear to be rescaled versions of cluster haloes \citep{moore1999}. Moreover haloes contain subhaloes whose properties such as the density profile are similar to those of haloes (e.g.~\citealp{diemand2008}; and \citealp{springel2008}). The subhalo mass and circular velocity distributions show little scale dependence if normalized in terms of the host halo mass or maximum velocity (e.g.~\citealp{kravtsov2004}; \citealp{gao2004}; and \citealp{reed2005}). Furthermore, the subhalo mass distribution is characterized by a power-law, $dN/d\ln M_{sub}\propto M_{sub}^{\alpha}$ with $\alpha=-0.8$~to~$-1.0$ (e.g.~\citealp{stoehr2003}; \citealp{gao2004}; \citealp{diemand2007}; \citealp{madau2008}; and \citealp{springel2008}). Note that a power-law index of -1 implies a scale-free distribution with a constant mass in substructure per logarithmic mass interval. The most recent  studies seem to favour $\alpha=-0.9$ for galactic haloes down to a subhalo mass of $\sim10^5\Msun$ (\citealp{madau2008} and \citealp{springel2008}). Substructure is not entirely independent of environment as the amount of it in a given halo depends on the halo's formation time or peak height and concentration; haloes that form earlier from high-$\sigma$ peaks have more substructure than low-$\sigma$ haloes of similar mass while the more concentrated the halo's density profile the more substructure it contains (e.g.~\citealp{bullock2001}; \citealp{gao2004}; and \citealp{zentner2005}).

\par
These ideas suggest a simple picture of substructure: haloes contain a scale-free distribution of subhaloes, which in turn contain a similar distribution of subsubhaloes, all the way down the CDM hierarchy. The scale of the bottom of the hierarchy is set by fundamental physics, namely the free-streaming and collisional dampening scales of dark matter. If, for example, dark matter is the neutralino, a Weakly Interacting Massive Particle (WIMP) predicted by supersymmetric extensions to the Standard Model (SUSY), the first objects form at a redshift of $z\gtrsim 60$ and have masses $M\lesssim10^{-6}\Msun$ \citep{green2004,green2005}. \citet{moorediemand2005} investigated the formation of these objects and postulated that there would be $\sim10^{15}$ of them in the Milky Way (MW) halo today. The formation of a rare, high redshift $0.014\Msun$ object was simulated by \cite{diemand2006}. They found that the subhalo mass function is very similar to that of cluster haloes. Again the suggestion is that self-similarity in halo's substructure extends all the way down the hierarchy.

\par
Self-similarity in the distribution of substructure may have important implications for a variety of experiments such as GLAST, which will search for $\gamma$-rays from DM self-annihilation. The annihilation signal is very sensitive to the slope of the subhalo mass distribution and the number of WIMP-scale subhaloes since substructure can boost the flux by factors of a 3-100 (e.g.~\citealp{diemand2007}; \citealp{kuhlen2008}; \citealp{pieri2008}; and \citealp{strigari2008}). The amount of dark matter locked up in substructure also has ramifications for direct DM detection as it reduces the density of the smooth background component of the Galactic halo but also increases the local density inside subhaloes \citep{kamionkowski2008}.

\par
The extrapolation used to make predictions of the $\gamma$-ray flux are non-trivial since the MW halo is a factor of $10^{18}$ times more massive than the smallest subhaloes. The estimate by \citet{moorediemand2005} of $\sim10^{15}$ small subhaloes neglects mergers and tidal interactions inherent in the hierarchical structure formation scenario. Haloes must survive similar-mass mergers and accretion, along with other dynamical processes that exist in galaxies \citep{zhao2007}. Furthermore, the choice of $\alpha$ is crucial as small changes in $\alpha$ by as little as $0.1$ can change the number of subhaloes at the bottom of the hierarchy by an order of magnitude or more.

\par
These results suggest that the internal properties of individual haloes (e.g.~the subhalo mass function) have no memory of the initial power spectrum. However, this would be suprising considering other internal properties of dark matter haloes, such as concentration of the density profile, depend on properties of the primordial power spectrum (e.g.~\citealp{reedprofiles2005}). For example, scale-free simulations by \citet{knollmann2008} show that haloes have less concentrated density profiles as $n\rightarrow-3$.

\par
There are other theoretical reasons to suspect that substructure is qualitatively different at small scales. As one approaches the bottom of the CDM hierarchy, $\neff$ monotonically decreases to $-3$ down to the cutoff at the WIMP free-streaming scale. The dimensionless power spectrum, $\Delta^2(k)\propto k^{\neff+3}$, becomes scale independent as $n\rightarrow-3$ and objects collapse simultaneously over a wide range in scales above the free-streaming scale. Haloes at these scales do not form in a clean hierarchical fashion and may not virialize before merging with or being accreted by other haloes, which may influence the distribution of substructure. In short, structure formation at the smallest scales in the CDM hierarchy may be qualitatively different from structure formation via hierarchical clustering that occurs at galactic scales.

\par
There are hints in previous studies that substructure does have a spectral dependence. \cite{diemand2006} showed substructure in a high redshift subsolar mass object was more susceptible to tidal disruption, with as much as $20-40\%$ of it being disrupted within an expansion factor of 1.3 as compared to $1\%$ in a low redshift cluster halo. \cite{springel2008} found that subgalactic subhaloes have less substructure than galactic haloes when regions of the same overdensity are compared. They also presented evidence that suggests the slope of the subsubhalo mass function flattens as the subhalo host mass decreases. The scale-free simulations of \cite{reed2005} appear to show that the subhalo velocity distribution flattens when $n<-2$.

\par
However, simulations of $n<-2$ cosmologies are notoriously difficult to perform due to finite volume effects \citep{smith2003}. Missing power from modes larger than the simulation box affect statistical quantities such as the two-point correlation function and halo mass function (\citealp{baglaray2005}; \citealp{power2006}; and \citealp{baglaprasad2006}). Provided large-scale modes have negligible amplitudes and the scale of interest is a small fraction of the box size, the internal properties of haloes appear unaffected. One can also correct for deviations in statistical quantities, though these corrections become increasingly large as $n\rightarrow-3$. Generally, it is preferable to simply halt a simulation before finite volume effects become an issue. We revisit the criterion presented by \cite{smith2003} that ensures the finite volume effects are negligible and show that it has {\em not} been met in the prior simulations of scale-free power spectra with $n<-2$.

\par
We run a pair of scale-free Einstein-de Sitter simulations in order to quantify the spectral dependence of the subhalo mass function. As most studies focus on clusters and galaxies, where $\neff\approx-1.8$ and $-2.1$ respectively, we choose spectral indices of $n=-1$ and $n=-2.5$ to bracket these scales in the CDM hierarchy. Finite volume effects are carefully considered in choosing the end point of our simulations. We search  all large haloes for subhaloes and vary the parameters of the group finding algorithm in order to quantify systematics. Our results show that subhalo mass function does depend on $n$, in contrast to previous results, and is sensitive to the parameters used to find subhaloes.

\par
Our paper is organized as follows: In section \Secref{sec:numreq}, we discuss particle number requirements and the difficulties with simulating $n\rightarrow-3$ cosmologies. The initial conditions and cosmological parameters used along with the technical details of the simulations are presented in \Secref{sec:methods}. In \Secref{sec:haloes}, we compare the halo mass distribution from our simulations with theoretical models. In \Secref{sec:subhaloes}, we present our results for the properties of subhaloes and the dependence of subhaloes on the spectral index. The paper concludes in \Secref{sec:discussion} with a summary and discussion.

\section{Particle requirements}
\label{sec:numreq}
\begin{figure}
    \centering
    \includegraphics[width=0.44\textwidth]{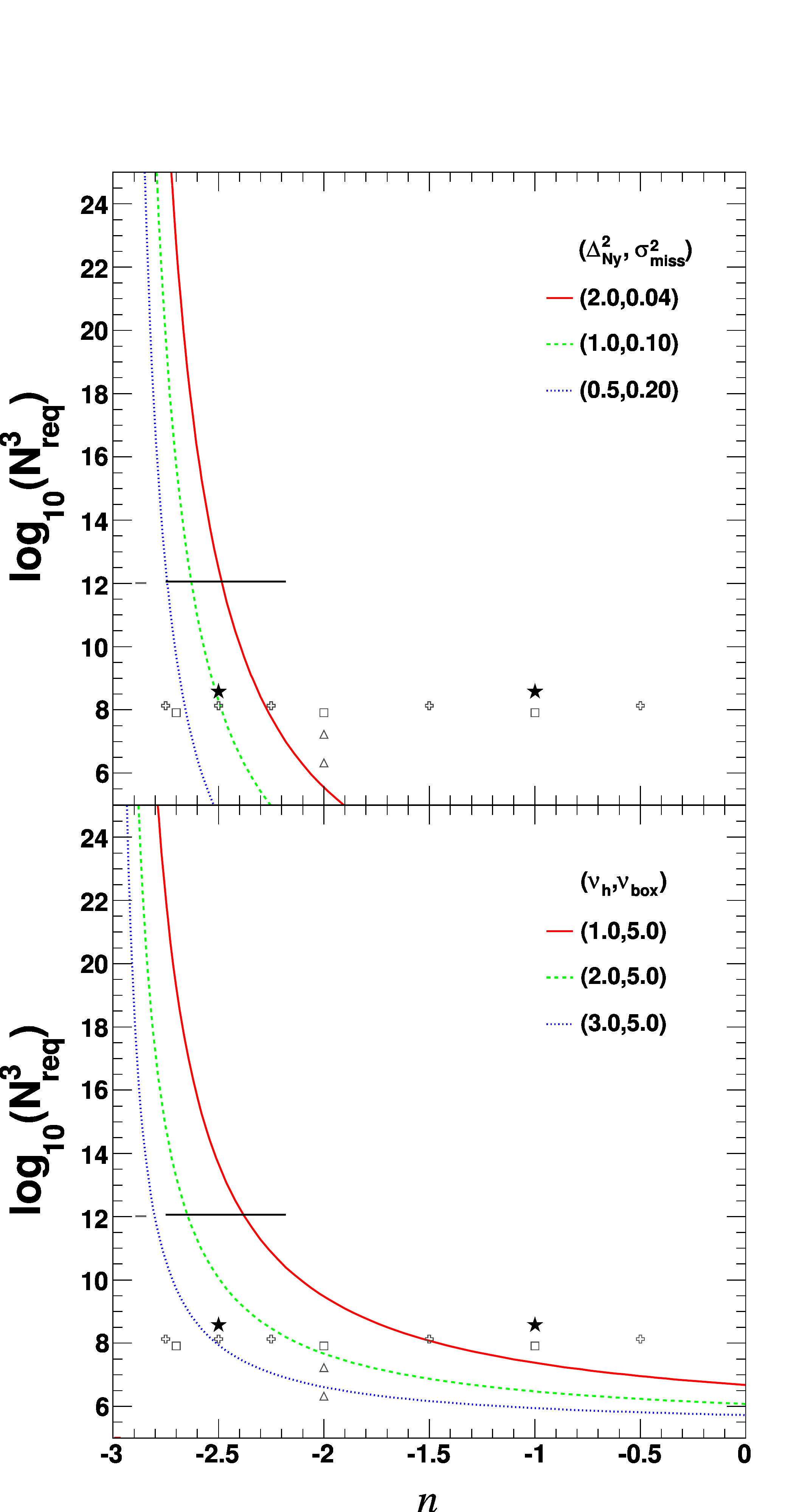}
    \caption{Number of particles required, $N^3_{\text{req}}$ versus spectral index $n$ based on \Eqref{eqn:nreqvar} (top) and \Eqref{eqn:nreqmass} (bottom) for a variety of constraints on the linearity large-scale modes and the nonlinearity of modes at the Nyquist scale. The constraints for the top panel are $(\Delta^2(k_{\text{Ny}}),\sigma^2_{\text{miss}})=[(2,0.04),(1.0,0.1),(0.5,0.2)]$ in solid, dashed and dotted curves respectively. The bottom panel indicates the requirements for haloes of $10^5$ particles to be $1\sigma,~2\sigma$~and~$3\sigma$ peaks in solid, dashed and dotted curves respectively with the box mass being a $5\sigma$ peak. Filled stars indicate the resolution and spectral index of the two simulations we ran. For comparison, also shown are scale-free simulations from \citet{knollmann2008} in open crosses, \citet{reed2005} in  open squares, and \citet{bertschinger1998} in open triangles. The effective resolution and effective spectral index of the CDM power spectrum from the Nyquist scale up to $k_{\text{halo}}=(4\pi\rho_{\text{bg}}/3M)^{1/3}$ for \citet{diemand2006,diemand2007} are shown in thin and thick horizontal lines respectively.}
    \label{fig:Nreq}
\end{figure}

Simulations are a compromise between two competing goals: good statistics, which favours the use of a physically large box, and high resolution, which promotes the use of a small box to achieve high resolution of individual objects. Since substructure is sensitive to the softening length used in simulations the second of these issues tends to dominate the choice of box size. Indeed, substructure in haloes composed of $\lesssim10^5$ particles tends to evaporates due to numerical softening effects (e.g.~\citealp{moore1999} and \citealp{klypin1999}).

\par
Choosing between a small box with a resolution high enough to resolve small-scale structure while still keeping large-scale modes in the linear regime is problematic. Simulations must be halted before the large-scale modes are nonlinear so as to ensure that mode-mode coupling is correctly represented, and that the amount of power due to missing large-scale modes is negligible. The dimensionless power spectrum is
\begin{equation}
    \Delta^2(k)=\frac{V}{(2\pi)^3}4\pi k^3 P(k),
\end{equation}
where $V$ is the normalization volume and $P(k)$ is the power spectrum. In what follows, $P(k)$ and $\Delta^2(k)$ refer to the power spectrum evolved according to linear theory. We assume a scale-free initial power spectrum so that
\begin{equation}
    P(k)=a^2Ak^n,\label{eqn:pk}
\end{equation}
where $A$ is the amplitude and $a$ is the cosmological scale factor. Modes are considered to be linear if $\Delta^2(k)\ll 1$ and nonlinear if $\Delta^2(k)\geq 1$ with the nonlinear scale defined by the relation $\Delta^2(k_{\text{NL}})=1$. The effective index of the power spectrum is
\begin{equation}
    \neff(k)\equiv\frac{d\ln\Delta^2}{d\ln k}-3~.
\end{equation}
For a simulation of size $L$ with $N^3$ particles, the power at a given $k$ is related to the power at the Nyquist wavenumber, $k_{\text{Ny}}=\pi N/L$, by
\begin{equation}
    \Delta^2(k)=\Delta^2(k_{\text{Ny}})\biggl(\frac{k}{k_{\text{Ny}}}\biggr)^{3+n}.
\end{equation}
Thus, the power at the box scale $k_{\text{b}}=2\pi/L$ is
\begin{equation} 
\Delta^2(k_{\text{b}})=\Delta^2(k_{\text{Ny}})\biggl(\frac{2}{N}\biggr)^{3+n}.
\label{eqn:kbox}
\end{equation}
We can use this definition to define an end point for a given simulation by requiring that the mode at the box scale is still linear.

\par
The impact of modes larger than the box can be quantified by considering missing variance, that is the difference between the variance integral for an infinite universe and the finite sum over modes within the box. \cite{smith2003} show that the missing variance for scale-free power spectra is well approximated by
\begin{equation}
    \sigma^2_{\text{miss}}=\frac{\Delta^2(k_b)}{3+n}F(3+n),
    \label{eqn:missvar}
\end{equation}
where $F(x)=1-0.31x+0.015x^2+0.00133x^3$ for $-3\leq n\leq1$. As $n\rightarrow-3$, $\sigma^2_{\text{miss}}\rightarrow\infty$, that is, missing power plays an increasingly important role. Ideally, one wants $\sigma^2_{\text{miss}}\rightarrow0$, though so long as $\sigma^2_{\text{miss}}\ll1$ the simulation will not suffer significantly from finite volume effects and the large-scale modes will still be linear. \citet{smith2003} use $\sigma^2_{\text{miss}}<0.04$ in their study of the nonlinear evolution of the power spectrum, which we also use as our strictest limit. However, there is not a precise value of $\sigma^2_{\text{miss}}$ that guarantees some level of accuracy will be achieved for any given realization.

\par
We can combine a desired nonlinearity at a given scale and our constraint on $\sigma^2_{\text{miss}}$ to obtain a minimum requirement on the number of particles used in a simulation. The minimum number of particles required to achieve a level of nonlinearity at the Nyquist scale, $\Delta^2(k_{\text{Ny}})$ is
\begin{equation}
    N^3_{\text{req}} =
    2^3\biggl[\frac{\Delta^2(k_{\text{Ny}})}{\sigma^2_{\text{miss}}}\frac{F(3+n)}{3+n}\biggr]^{3/(3+n)}.
    \label{eqn:nreqvar}
\end{equation}
A useful choice is $\Delta^2(k_{\text{Ny}})>1$ since this ensures that the simulation is in the nonlinear regime. It is evident from this equation that for $\Delta^2(k_{\text{Ny}})/\sigma^2_{\text{miss}}>1$, the minimum number of particles rises super-exponentially as $n\rightarrow-3$.

\par
Another measure of nonlinearity is the mass variance, $\sigma^2(M)$, given by
\begin{equation}
    \sigma^2(M)=\frac{V}{(2\pi)^3}\int P(k)|\hat{W}^2(kR)|d^3{\bf k},
\end{equation}
where $\hat{W}(x)=(3/x^3)(\sin x - x\cos x)$ is the Fourier transform of the top-hat window function, $R=(3M/4\pi\rho_{\text{bg}})^{1/3}$ and $\rho_{\text{bg}}$ is the background mass density. A mass scale is linear if $\sigma(M)<\delta_{\text{sc}}$, where $\delta_{\text{sc}}$ is the critical density for spherical collapse, and nonlinear if $\sigma(M)>\delta_{\text{sc}}$. The characteristic mass scale $M_*$ is defined by the relation $\sigma(M_*)=\delta_{\text{sc}}$. We require $\sigma(M_{\text{box}})\ll\delta_{\text{sc}}$, where $M_{\text{box}}=\pi N^3/6$ is the mass enclosed in the largest sphere contain in the simulation volume. As $\sigma(M)$ is a statistical quantity, the criteria that $\sigma(M_{\text{box}})=\delta_{\text{sc}}/\nu$, where where $\nu$ is the height of density field in rms units, is a probabilistic one. For example, with $\nu=2$ and Gaussian initial conditions, $4.5\%$ of an ensemble of simulations would be nonlinear at the box scale. We choose $\nu_{\text{box}}\sim5$, ensuring that the probability that the box mode is nonlinear is $\sim10^{-6}$. Note that the effective index at a given mass scale is
\begin{equation}
    \neff(M)=-3\frac{d\ln\sigma^2(M)}{d\ln M}-3.
\end{equation}

\par
Using the mass variance, one can impose evolutionary criteria to calculate the minimum number of particles required.  Generally, studies focus on haloes of a particular mass scale, $M$, which correspond to density peaks above some threshold $\nu$. The minimum number of particles required to investigate haloes originating from $\nu_{h}\sigma$ peaks of composed of $N_{h}$ particles is:
\begin{equation}
    N^3_{\text{req}} =
    \frac{6N_h}{\pi}\biggl(\frac{\nu_{\text{box}}}{\nu_{h}}\biggr)^{6/(3+n)}.
    \label{eqn:nreqmass}
\end{equation}
As $n\rightarrow-3$, the mass variance becomes independent of scale and the minimum number of particles again rises super-exponentially.

\par
We demonstrate the relative importance of these constraints in \Figref{fig:Nreq} by plotting $N^3_{\text{req}}$ as a function of $n$. The curves indicate the minimum number required for a given set of constraints, with the constraints becoming relaxed as one goes from solid to dashed to dotted. To be representative of a given cosmology and spectral index, simulations must lie above these curves. 
Note that the number of particles needed to attain a highly evolved simulation $\Delta^2(k_{\text{Ny}})\gtrsim1$ while still limiting $\sigma^2_{\text{miss}}\leq0.04$ for $n<-2$ becomes increasingly impractical. For very negative indices, a more reasonable goal is evolving the simulation till the Nyquist scale is just nonlinear while $\sigma^2_{\text{miss}}\sim0.10$. Again, we stress that though our choices of $\sigma^2_{\text{miss}}$ and $\nu_{\text{box}}$ are meant to be conservative, these criteria do not guarantee that our study is free of finite volume effects. Simulations are always missing power and this missing power, even if small, can subtly affect the evolution of the system as shown in \cite{powerpaper}. A rigorous examination of finite volume effects on the subhalo population will require further study.

\par
The missing variance is not as large in $\Lambda$CDM as it is for scale-free power spectra due to the turnover in the power spectrum but still goes as $\Delta^2(k_{\text{b}})$. This figure clearly shows the difficulties with modelling the bottom of the CDM hierarchy where $\neff\approx-2.8$. The highest resolution simulation of SUSY haloes by \cite{diemand2006} has $\neff\approx-2.85~$to~$-2.75$. As they examine the substructure of a very rare $3.5\sigma$ peak and halt their simulation early enough, the missing variance does not exceed $\approx0.05$. However, any extrapolations based on a $3.5\sigma$ peak should be treated with caution.

\par
Equations (\ref{eqn:nreqvar}) and (\ref{eqn:nreqmass}) can also be rearranged to determine when to halt a simulation for a given $\sigma^2_{\text{miss}}$ or  $\sigma^2(M_{\text{box}})$. Since we are interested in substructure, we focus on $\gtrsim10^5$ particles for $n=-2.5$ and $N^3=720^3$, halting the simulation when $\sigma(M_{\text{box}})\approx\delta_{\text{sc}}/4.9$ means that these haloes originate from rare $\sim2.5\sigma$ peaks. If our $n=-1$ simulation is evolved to the same $\sigma(M_{\text{box}})$, the haloes composed of $10^5$ particles would originate from common $\sim0.4\sigma$ peaks. Using $\sigma^2(M)\propto a^2$, it is a simple matter to calculate how many e-foldings the simulation must be evolved to achieve this desired end state.
\section{Numerical methods}
\label{sec:methods}
\subsection{How to interpret scale-free simulations}
\label{sub:scalefree}
\begin{figure}
    \centering
    \includegraphics[width=0.44\textwidth]{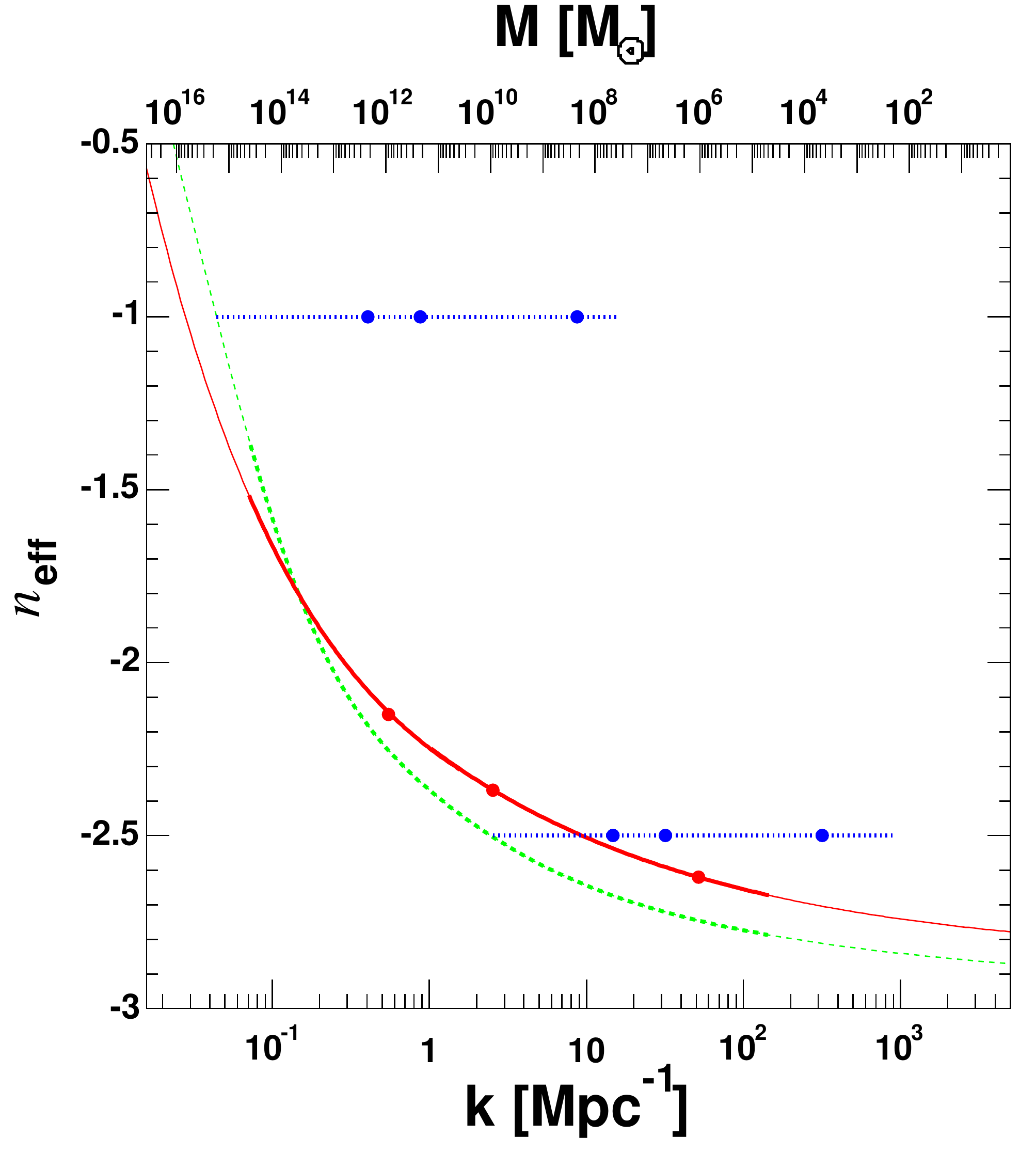}
    \caption{Effective spectral index of the CDM power spectrum versus mass (solid) and wavenumber (dashed). Highlighted in thick lines are the wavenumbers and mass scales sampled in simulations such as the Via Lactea simulation and the Aquarius Project. Various mass scales in the CDM hierarchy are indicated by filled circles, with the largest mass corresponds to the an MW halo ($10^{12}\Msun$) and the other two points indicating the largest and smallest galactic subhalo masses examined in previous studies, $10^{10}\Msun$ and $10^6\Msun$ respectively. Also shown are the indices of our $n=-1$ and $n=-2.5$ simulations (dotted) with the filled circles indicating masses of $5\times10^5$, $10^4$ and $100$ particles corresponding to the mass of haloes , largest subhaloes and smallest subhaloes examined.}
    \label{fig:viacomp}
\end{figure}
The varying slope of the $\Lambda$CDM power spectrum is key to interpreting scale-free simulations in the context of the CDM hierarchy. In \Figref{fig:viacomp}, we show the effective spectral index as a function of wavenumber and mass based on the $\Lambda$CDM power spectrum of \cite{eisenstein1998}. Highlighted are scales sampled in the ``Via Lactea'' (VL) simulation along with certain mass scales. Most numerical studies focus on clusters and galaxies, corresponding to $\neff\approx-1.8$~to~$-2.2$. Also shown are the indices of our simulations, where the scale of the box size is set to match the scale at which the CDM power spectrum has the same index. Our choice of indices is meant to bracket galaxy and cluster scales. In essence, haloes found in an $n=-2.5$ simulation are representative of high redshift haloes surrounding the first protogalaxies. Objects at this scale may also become the larger subhaloes found in galaxies. Though the actual power spectrum is steeper at these scales with $\neff\approx-2.7$, we satisfy ourselves with $n=-2.5$ as it is the steepest spectra we can reasonably simulate, though even at this index we are pushing the limits outlined in \Secref{sec:numreq}.

\subsection{Simulations}
\label{sub:sims}
\begin{table}
\centering
\caption{Summary of Simulations Run.}
\begin{tabular}{ccccc}
\hline
\hline
    $n$ & $N^3$ & $(a_f/a_i)$ & $a_f$ & Softening Length $\epsilon_s$ \\
\hline
    -1   & $720^3$ & 42.8 & 0.214 & 1/30(L/N)\\
    -2.5 & $720^3$ & 17.5 & 0.564 & 1/30(L/N)\\
\end{tabular}
\label{tab:sims}
\end{table}
We run scale-free Einstein-de Sitter $(\Omega_{\text{m}}=1$) cosmological simulations with power-law power spectra as shown in \Eqref{eqn:pk}. We choose an amplitude and initial scale factor so that the maximum initial displacement for any given particle is less than 1/2 the initial inter-particle spacing. Due to the random nature of the initial density field this choice does not correspond precisely to a specific amplitude requirement across all power spectra, but ensures that the simulation starts in the linear regime. 
We draw both simulations from the same random realization and normalize the amplitudes so that $\sigma(M=15\times10^{6}\text{ particles},a=1.0)=0.9$, which is equivalent to normalizing using $\sigma_8$. A final issue concerns transients in the density and velocity field generated by initial conditions generator. To minimize the effect of transients, we use initial conditions generated by second-order Lagrangian perturbation theory (2LPT) rather than the standard Zeldovich approximation (ZA) \citep{crocce2006b}. The simulations are run using the parallel N-body tree-PM code GADGET-2 \citep{gadget2}. We halt the $n=-2.5$ simulation when haloes composed of $\gtrsim10^5$ particles originate from $\gtrsim2.5\sigma$ peaks and the box scale corresponds to a rare $4.9\sigma$. Evolving farther in order to form larger haloes and further reduce numerical softening effects is not possible since, even at the our chosen end point, we are pushing the limits with $\sigma^2_{\text{miss}}\approx0.10$. We also analyze the $n=-1$ simulation when $\gtrsim2.5\sigma$ haloes are composed of $\gtrsim 10^5$ particles for the sake of consistency, though $\sigma^2_{\text{miss}}\approx0.001$ at this point. A summary of the simulations is given in \Tableref{tab:sims}.

\section{Haloes}
\label{sec:haloes}
\begin{figure*}
    \centering
    \includegraphics[width=0.98\textwidth]{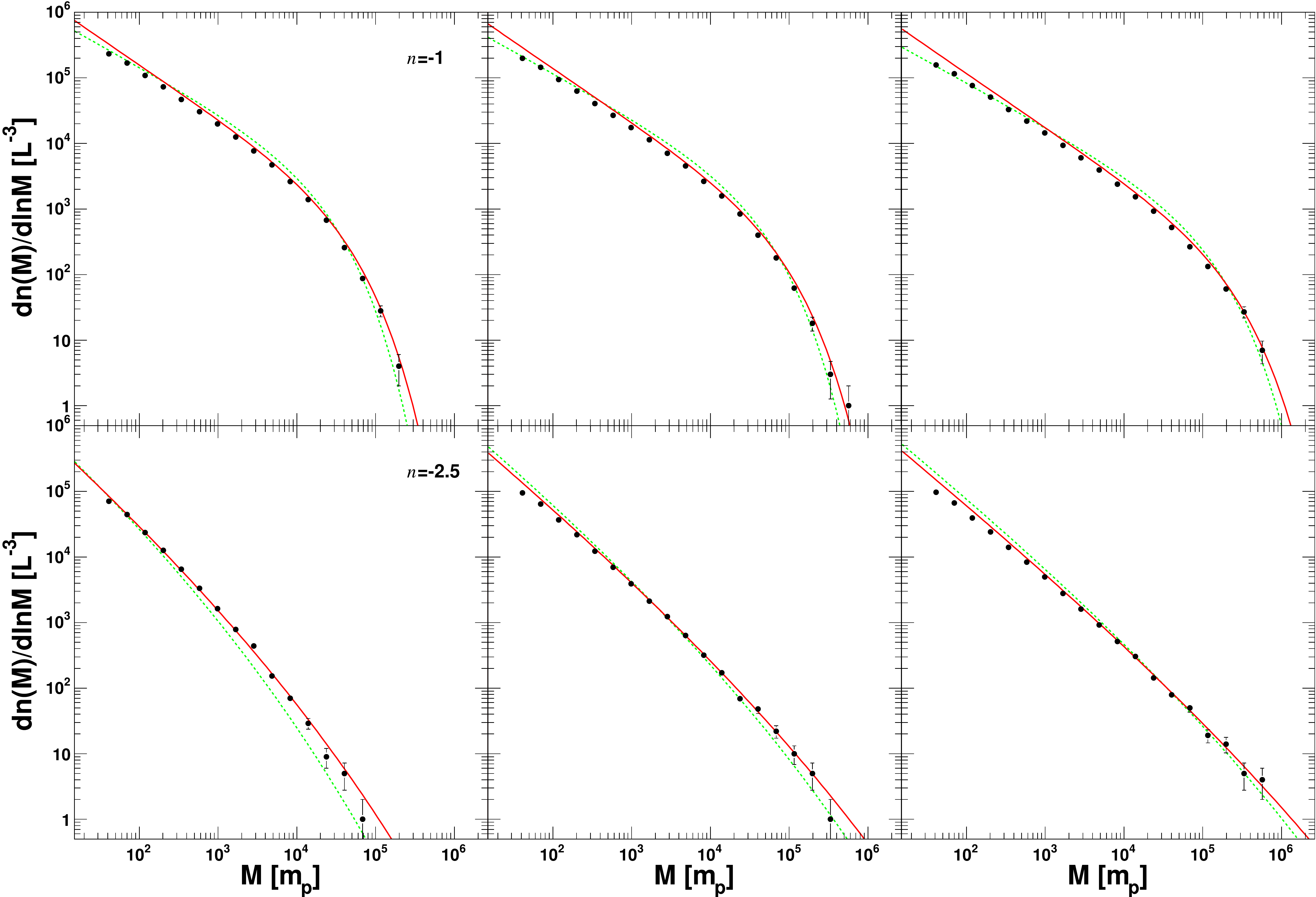}
    \caption{The co-moving halo number density of mass $M$, $dn(M,z)/d\ln M$ denoted by filled points with $1\sigma$ error bars at $n=-1$  (top) and $n=-2.5$ (bottom). Curves are predictions from analytic fitting functions proposed by ST (solid) and PS (dashed). Going from right to left corresponds to going to progressively lower redshifts with right panel corresponding to the final redshift analyzed.}
    \label{fig:halomass}
\end{figure*}

We identify haloes at each time step using a friends-of-friends (FOF) group finder with a linking length $\ell_{h}=0.2$ times the comoving interparticle spacing $\Delta x=L/N$ \citep{fof}. Only FOF groups with more than 32 particles are kept in the halo catalogue. The comoving halo number density is shown in \Figref{fig:halomass} at three different redshifts along with theoretical predictions from \cite{shethtormen1999} (ST) and \cite{pressschechter} (PS). The \citetalias{pressschechter} mass function is based on the spherical collapse model while the \citetalias{shethtormen1999} incorporates ellipsoidal collapse with additional mass function parameters calibrated using large-scale $\Lambda$CDM simulations. For both simulations, the \citetalias{shethtormen1999} mass distribution is in better agreement than \citetalias{pressschechter} for all redshifts. 

\section{Substructure}
\label{sec:subhaloes}
\subsection{Searching for Subhaloes}
\label{sec:subhalosearch}
Identifying subhaloes is more difficult than identifying haloes. The key issue with defining the outer boundary of a subhalo embedded in a gravitationally bound object. Several methods have been developed to search for substructure, including SKID \citep{skid} and SUBFIND \citep{subfind}. In this paper, we use a phase-space FOF (6DFOF) algorithm, which is similar to the one described in \cite{diemand2006}. We limit our subhalo search to haloes originating from $\gtrsim2.5\sigma$ peaks, constraining our analysis to 22 haloes composed of $\gtrsim3.2\times10^5$ particles and 29 haloes composed of $\gtrsim1.2\times10^5$ particles for the $n=-1$ and $n=-2.5$ simulation respectively. Each halo is associated with a velocity scale $\Delta v=(GM_h/R)^{1/2}$ where $R^3=M_h/(4\pi\rho_{\text{bg}}\ell_{h}^{-3}/3)$, $M_h$ is the mass of the halo found using a linking length of $\ell_h$ and $\rho_{\text{bg}}$ is the background density. Two particles with phase-space coordinates $({\bf x}_1,{\bf v}_1)$ and $({\bf x}_2,{\bf v}_2)$ are linked if
\begin{equation}
    \frac{({\bf x}_1-{\bf x}_2)^2}{(\ell_s\Delta x)^2}+\frac{({\bf v}_1-{\bf v}_2)^2}{(b_v\Delta v)^2}<1,
\end{equation}
where $\ell_s$ and $b_v$ are the dimensionless parameters defining the physical and velocity linking lengths respectively. We pass candidate subhaloes through an unbinding routine that checks whether an object is self-bound and removes any unbound particles \citep{springel2008}. Only subhaloes with 20 particles or more are kept in the catalogue. This unbinding routine is a necessary but time consuming process that removes unbound particles from subhalo candidates and also eliminates spuriously linked particles. The fraction of unbound candidates in the initial 6DFOF catalogue varies slightly with $b_v$ and is $\approx0.05$ and $\approx0.70$ for the $n=-1$ and $n=-2.5$ simulation respectively.

\par
We set $\ell_s=0.10$, thus limiting our search to regions within haloes with physical overdensities of $\rho/\rho_{\text{bg}}\gtrsim1000$, conversely we try several velocity linking lengths, $b_v=[0.0125,0.025,0.05,0.10]$. For an isolated spherical overdensity, increasing $b_v$ amounts to placing a higher circular velocity cutoff in defining the boundary of a candidate subhalo. The mass associated with a phase-space peak would continuously increase as $b_v$ is increased were it not for our unbinding routine which effectively imposes a tidal limit. In such a case, our technique is analogues to that used by \cite{diemand2007}. They found phase-space peaks using the 6DFOF algorithm and then estimated a tidal mass by assuming spherical symmetry and fitting an \citetalias{nfw} profile plus background to circular velocity profiles of the peaks. Since our method does not impose a particular density profile, we can indirectly examine the validity of spherical symmetry as $n\rightarrow-3$ and any systematic bias this assumption introduces in the resulting subhalo mass function by varying $b_v$.

\par
\begin{figure*}
    \centering
    \includegraphics[width=0.495\textwidth]{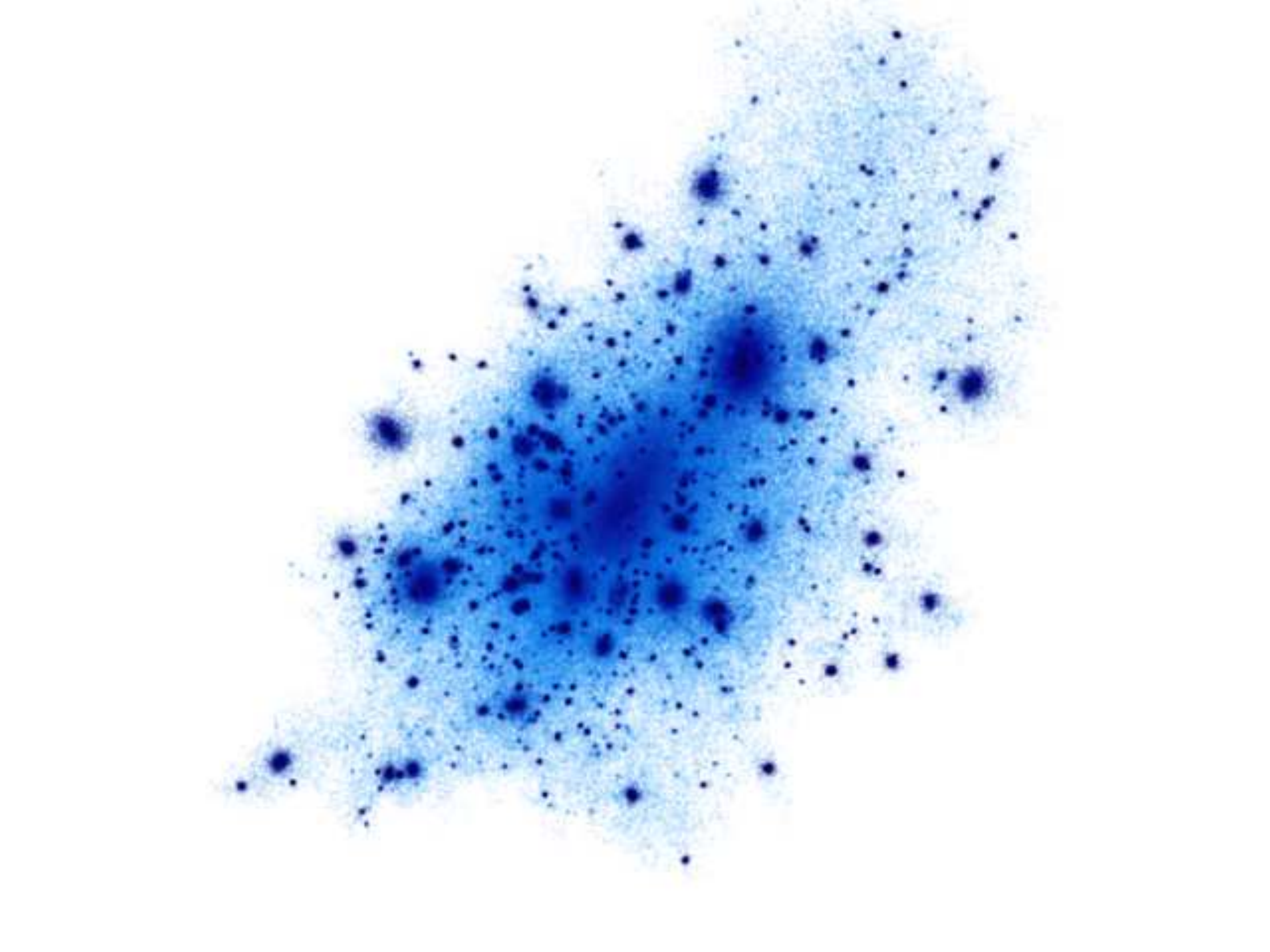}
    \includegraphics[width=0.495\textwidth]{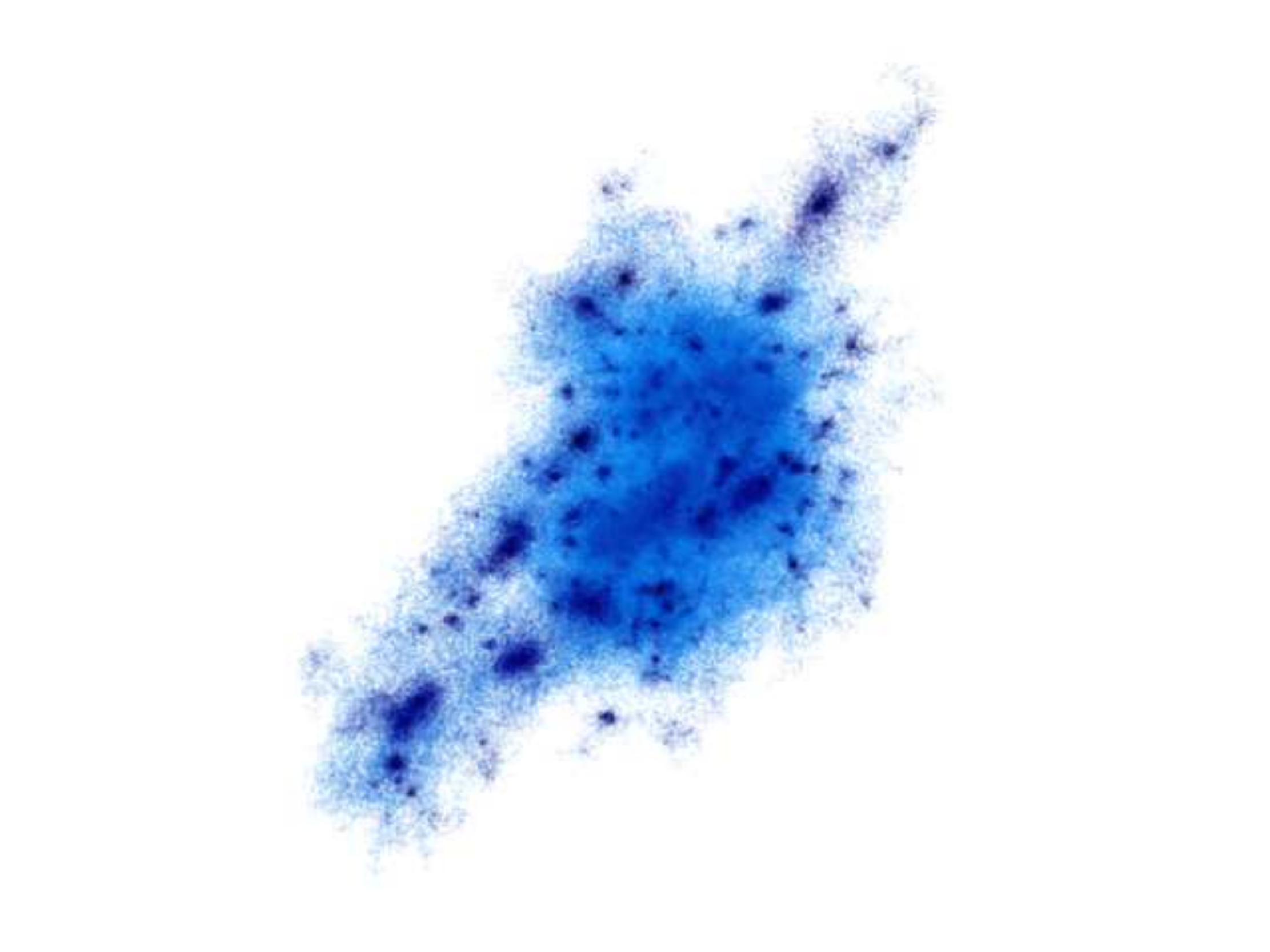}
    \includegraphics[width=0.495\textwidth]{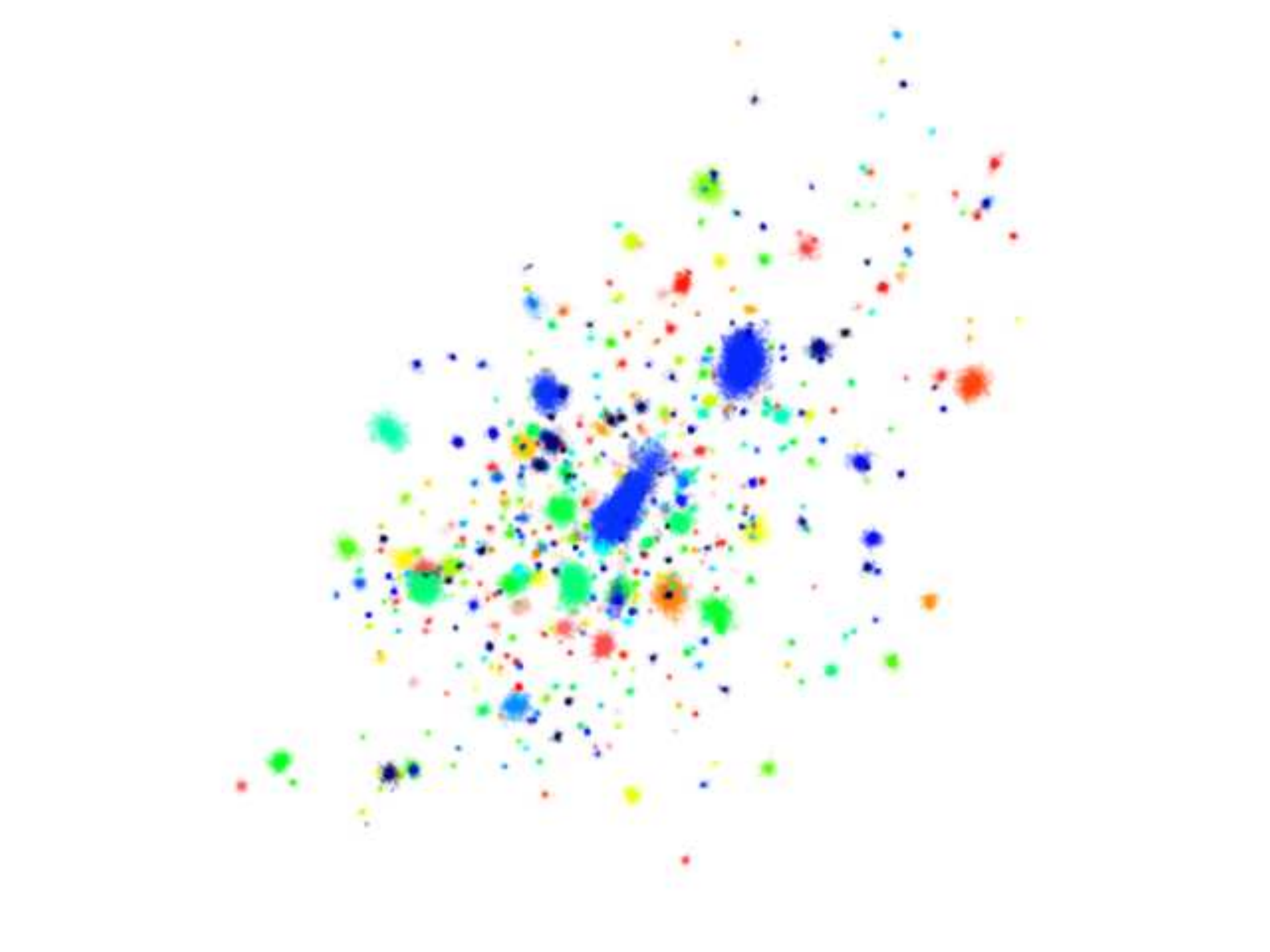}
    \includegraphics[width=0.495\textwidth]{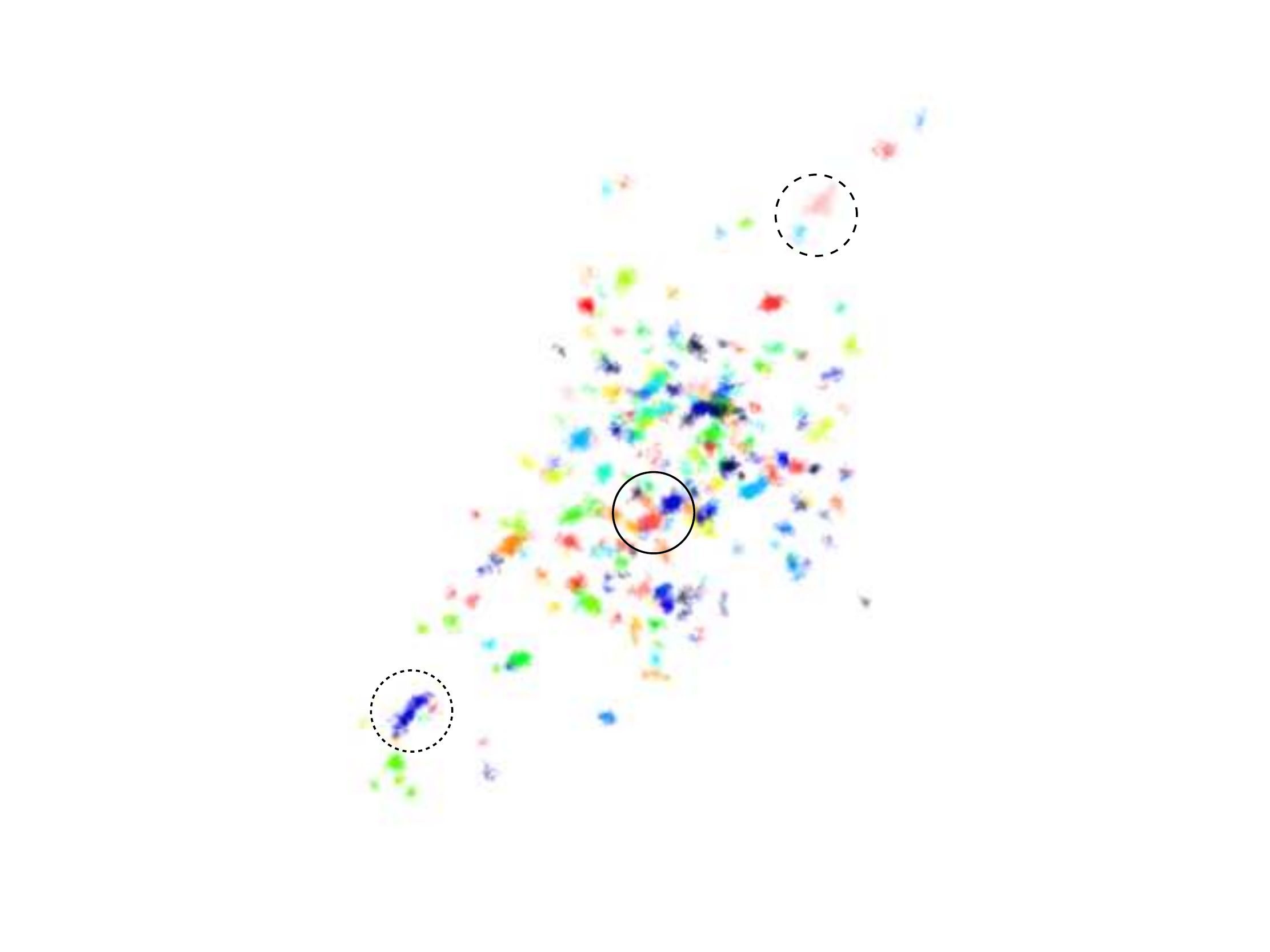}
    \includegraphics[width=0.495\textwidth]{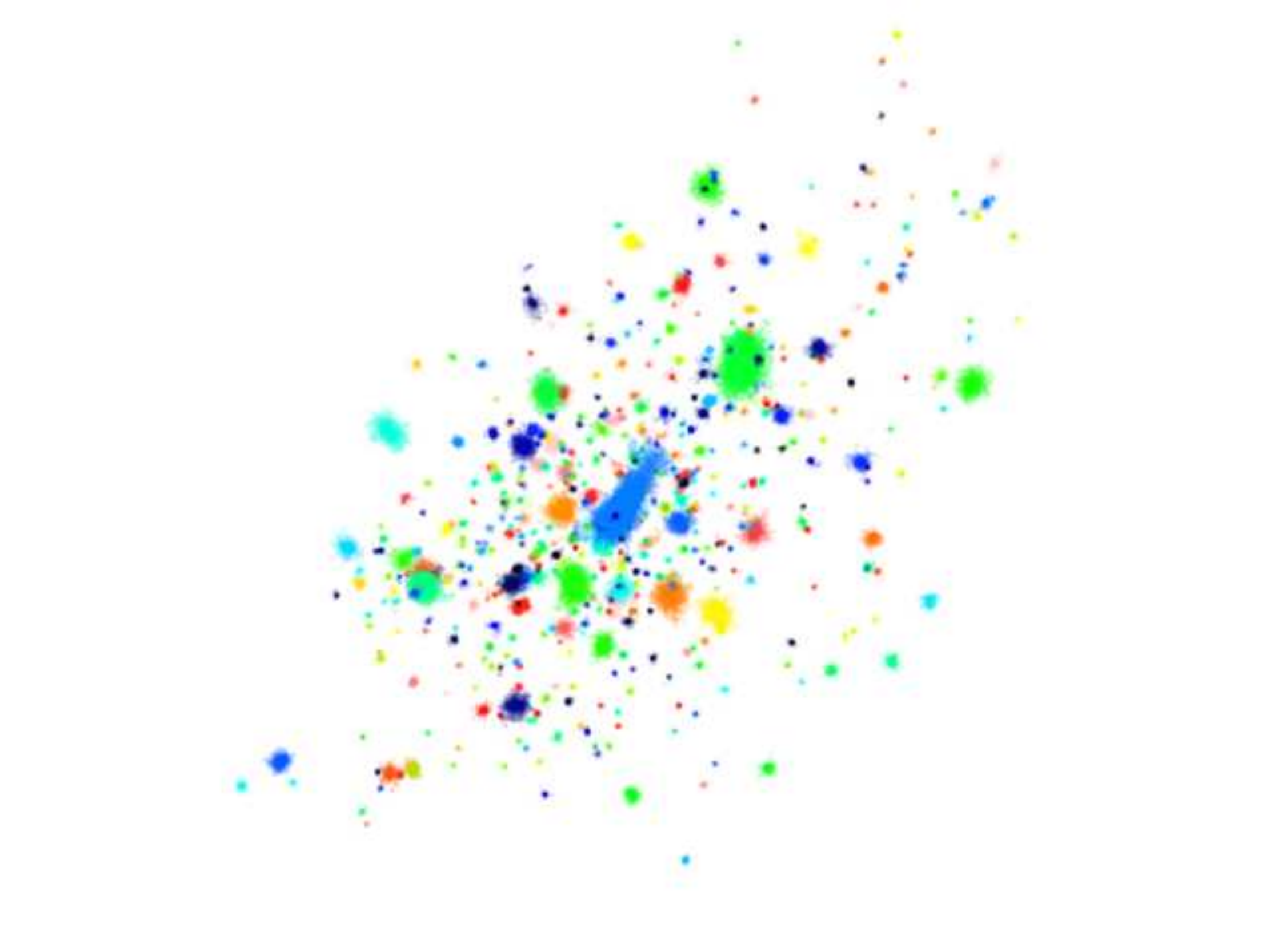}
    \includegraphics[width=0.495\textwidth]{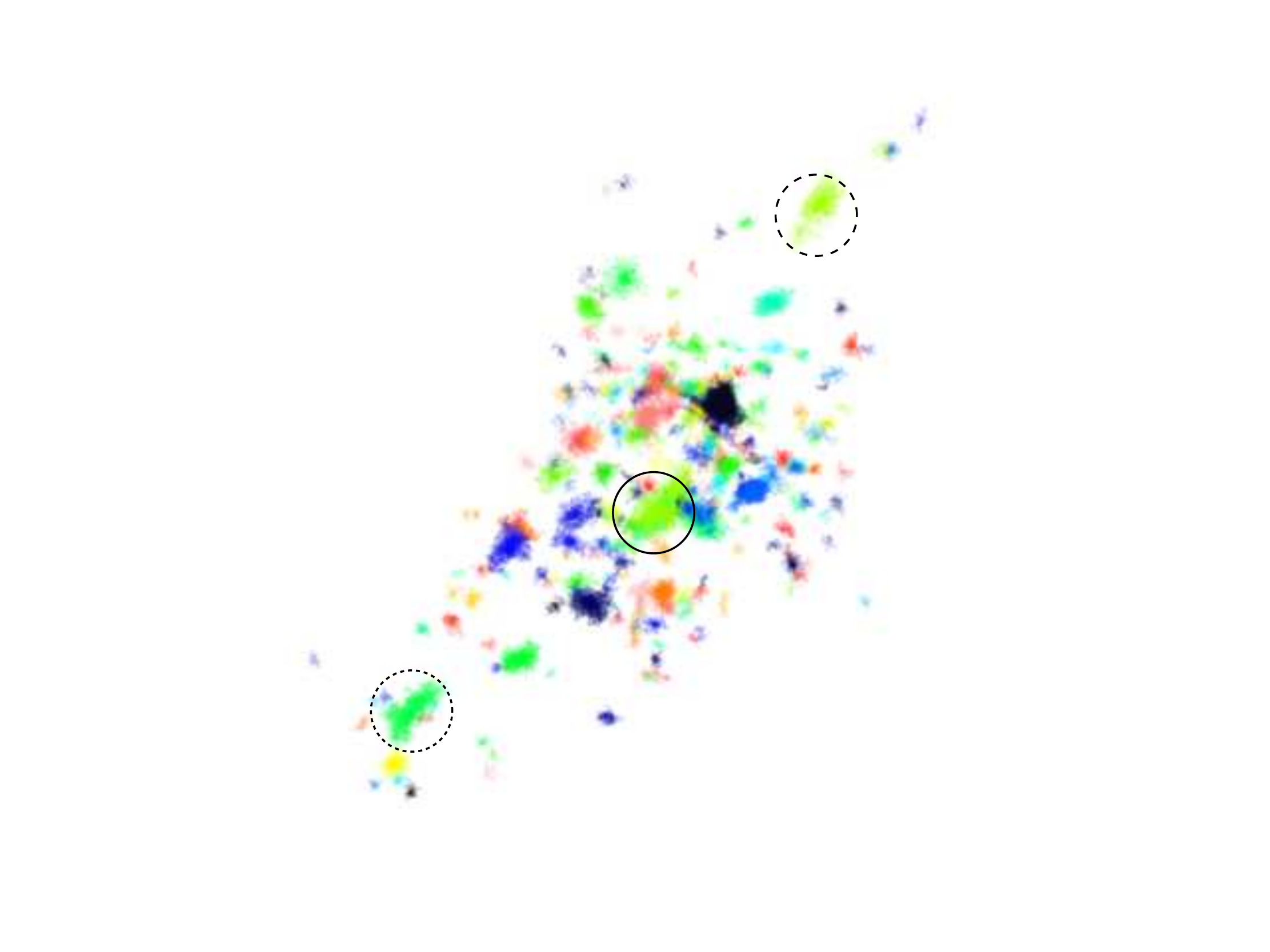}
    \caption{Two haloes originating from $\approx3\sigma$ peaks in the $n=-1$ (left) and $n=-2.5$ simulation (right) composed of $6.2\times10^5$ and $5.5\times10^5$ particles respectively. The first row shows the phase-space density of the haloes using a logarithmic colour scale, where dark blue regions indicate high phase-space density.In the next two rows, particles are colour coded according to group number found with 6DFOF. Shown are subhalo candidates found with $b_v=0.05$ and $0.0125$ in the middle and bottom rows, respectively. Three regions are circled in the $n=-2.5$ simulation to highlight the variation in particles linked by the 6DFOF algorithm with different $b_v$ (see discussion in text).}
    \label{fig:subcand}
\end{figure*}

In \Figref{fig:subcand}, we show different representations of two large haloes, one from each of our simulations. The two haloes originated from $\sim 3\sigma$ peaks and are composed of $\sim6\times10^5$ particles. For each halo we show the phase-space density calculated using EnBiD \citep{sharma2006} and the subhalo candidates found using two different velocity linking lengths. This figure demonstrates that there are significant substructure differences between the simulations. Substructure at $n=-2.5$ appears more triaxial and less dense than at $n=-1$ and, overall, the $n=-1$ halo looks very similar to a cluster or galaxy halo. At $n=-1$, substructure consists of distinct spherical overdensities with well defined boundaries, whereas substructure is less evident at more negative indices, (see \citealp{diemand2006}).
Of the 525 subhalo candidates found using $b_v=0.05$, 464 are bound and contain a total of $\approx20\%$ of the host halo's mass, though no single subhalo contains more than $5\%$ of the halo's mass. In contrast, the $n=-2.5$ halo has 285 candidates, of which only 21 are bound. These bound subhaloes contain $2\%$ of the halo's mass.

Comparing the impact of different velocity linking lengths $b_v$, we find that at $n=-2.5$, using $b_v\geq0.05$ appears to spuriously link phase-space peaks into long, unbound, filamentary structures, which are subsequently eliminated from the catalogue by our unbinding routine. This is the case in the central region outlined by a solid black circle. However, not all peaks are linked into artificial filaments. Occasionally, peaks are close enough to one another in phase-space and are linked by more than a few particles. Two candidate subhaloes, which are split at small $b_v$, are linked at larger $b_v$ into a bound triaxial object. An example of such is circled in a black dashed circle in \Figref{fig:subcand}. The dotted circle indicates a subhalo that is non-spherical regardless of $b_v$ used. These results visually demonstrate that the boundary of subhaloes appears less well defined as $n\rightarrow-3$.

\subsection{Phase-space structure of haloes}
\begin{figure*}
    \centering
    \includegraphics[width=0.98\textwidth]{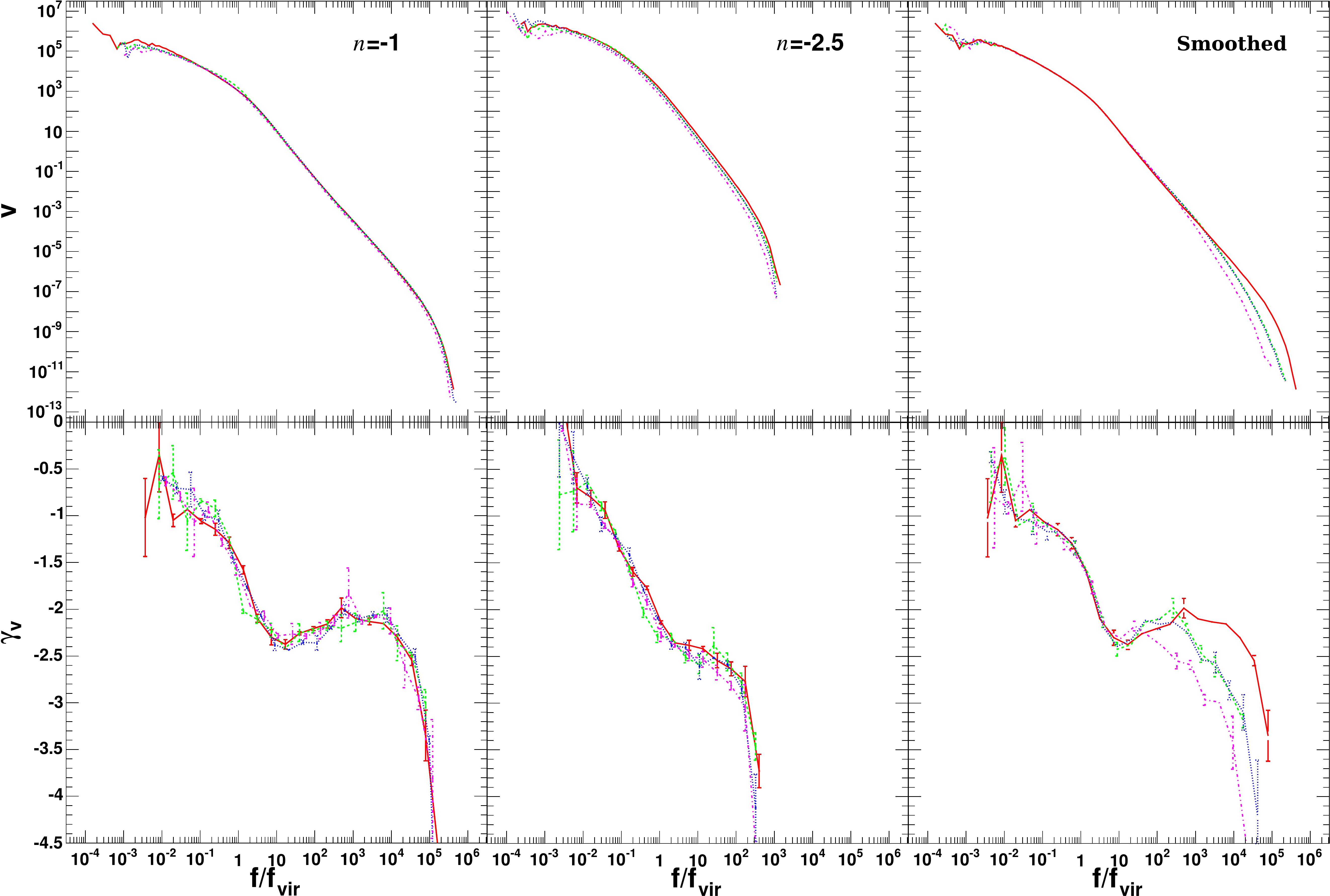}
    \caption{The normalized volume distribution function $v(f)$ (top) and logarithmic slope $\gamma_v$ (bottom) for 4 haloes consisting of $\gtrsim3\times10^5$ particles. The left and middle columns correspond to $n=-1$ and $n=-2.5$ simulation respectively. The solid lines corresponds to the haloes shown in \Figref{fig:subcand}, the other line types are ordered in decreasing halo mass going from dashed, dotted to dashed-dotted. The right column correspond to a $n=-1$ halo where the substructure has been smoothed with increasing smoothing going from dashed to dotted to dashed-dotted and the reference halo denoted by the solid curve.}
    \label{fig:halophase}
\end{figure*}

The upper panels of \Figref{fig:subcand} provide the phase-space density as a function of position. Of course the phase-space distribution $f({\bf x}, {\bf v})$, which provides a complete description of a dynamical system, is a function of six variables, making it cumbersome to deal with. As an alternative, we examine halo phase-space structure using the volume distribution function $v(f)$, which is the volume of phase-space occupied by phase-space elements of density $f$ in an interval $df$ \citep{arad2004}. In \Figref{fig:halophase}, we plot $v(f)$ (normalized so that $\int v(f)df=1$) along with the logarithmic slope, $\gamma_v\equiv d\ln v(f)/d \ln f$ for four haloes with $\gtrsim3\times10^5$ particles. Haloes from the $n=-1$ simulation have larger volumes with high phase-space density and span a greater range in $f$ than the haloes from the $n=-2.5$ simulation. At $n=-1$, $\gamma_{v}$ decreases to $\approx-2.5$ with increasing $f$ followed by a bump, and has a very similar form to that of $\Lambda$CDM galactic and cluster haloes examined by \cite{sharma2006}. The slope of haloes at $n=-2.5$ is shallower at low $f$ and plateaus around a slope of $-2.5$. Several authors have attempted to explain the form of $\gamma_v$ for $\Lambda$CDM haloes using a toy model where haloes are modelled by a Hernquist sphere that contains smaller Hernquist spheres representing subhaloes (e.g.~\citealp{arad2004}; \citealp{binney2005}; and \citealp{sharma2006}). \cite{sharma2006} found that the general shape of a $\Lambda$CDM galactic halo with substructure can be reproduced with these toy models and that the size of the bump appeared to be related to the amount of substructure.

\par
To examine whether the size of the bump in the slope is related to the phase-space density contrast between substructure and the background, we smooth out substructure in the $n=-1$ halo shown in \Figref{fig:subcand}. To smooth the halo we first calculate its morphology via the inertia tensor (see below) at different radii. Particles are then moved in a random direction through a small random angle on the surface of the ellipsoid determined by the mass distribution interior to their radii. This simple process leaves the radial density profile and overall morphology unchanged to within $\lesssim1\%$ while reducing the density contrast of substructure, thereby decreasing the number of candidates and bound subhaloes by a factor of $2-3$ and $8-60$ respectively. The result is shown in the right column of \Figref{fig:halophase}. We find that the size of the bump is influenced by both the amount and phase-space density contrast of substructure. The lack of a bump in the haloes from the $n=-2.5$ simulation shows that the contrast between substructure and the smooth background of a halo decreases as $n\rightarrow-3$. The differences in $\gamma_v$ do not arise solely due to differences in halo morphology as the haloes in either simulation have similar morphologies. Thus, the logarithmic slope of $v(f)$ can be a sensitive tool of the subhalo mass function and hence the differences seen in \Figref{fig:halophase} are suggestive that there may be a spectral dependence in both the amplitude and the slope of subhalo mass function.

\subsection{Morphology}
\label{sec:subhalomorph}
To determine the morphology of a subhalo we follow \cite{dubinski1991} and \cite{allgood2006} and diagonalize the weighted moments of inertia tensor
\begin{equation}
    \tilde{I}_{i,j}=\sum\limits_n \frac{x_{i,n} x_{j,n}}{r^2_{n}}.
\end{equation}
The ellipsoidal distance between the subhalo's centre of mass and the $n$th particle is
\begin{equation}
    r_n^2=x_n^2+(y_n/q)^2+(z_n/s)^2,
\end{equation}
where $q$ and $s$ are the intermediate-to-major and minor-to-major axis ratios respectively. The axis ratios are calculated after unbound particles are removed for subhaloes composed of $>100$ particles.

\par
\begin{figure*}
    \centering
    \includegraphics[width=0.61\textwidth]{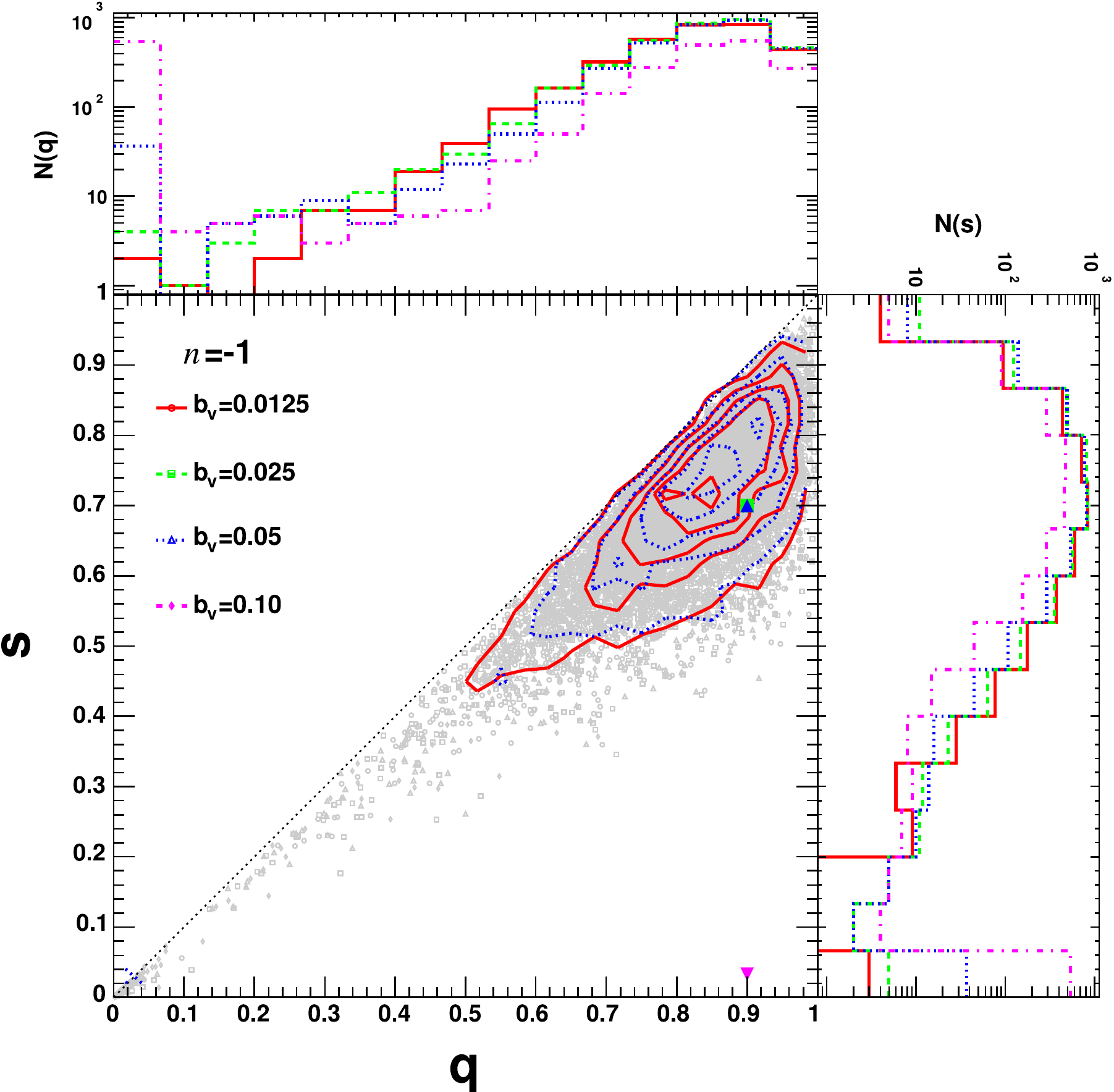}
    \includegraphics[width=0.61\textwidth]{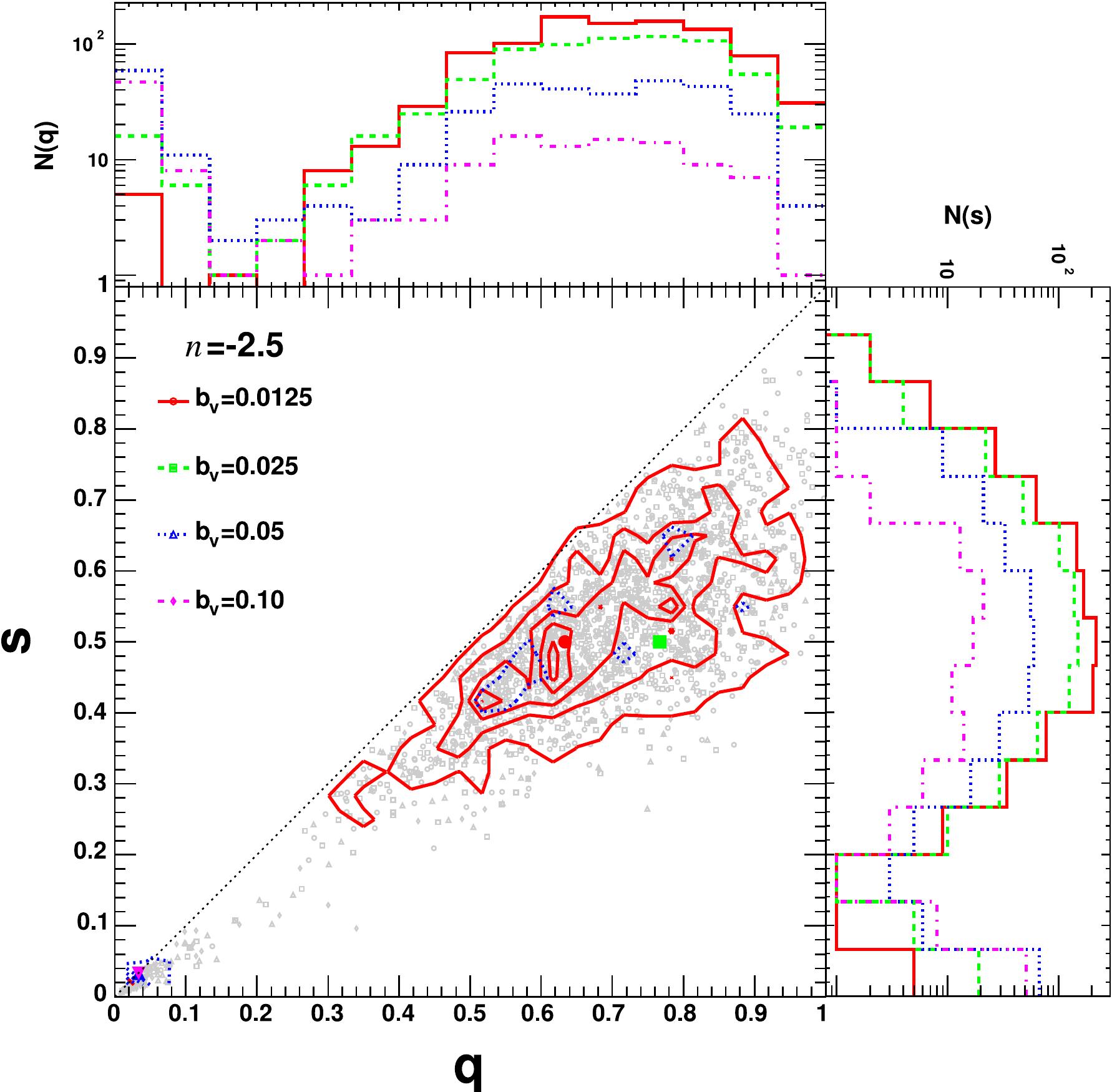}
    \caption{Scatter plot of major and minor axis ratios, $q$ and $s$, for subhaloes found in the $n=-1$ (top) and $n=-2.5$ (bottom) simulations. Also shown are the projects of these distributions. The grey open circles, squares, triangles and upside down triangles and solid, dashed, dotted, and dashed-dotted lines correspond to $b_v=0.0125, 0.025, 0.05$~and~$0.10$ respectively. For clarity we only show a random subsample $(10\%)$ and plot contours for $b_v=0.0125$ and $b_v=0.05$. Filled points indicate the peaks the histograms and follow the same marker scheme as the scatter plot. The thin dashed line corresponds to $q=s$.}
    \label{fig:submorph}
\end{figure*}

Figure \ref{fig:submorph} shows the distribution of subhaloes in terms of $q$ and $s$. We see that subhaloes are generally more triaxial at $n=-2.5$ compared to $n=-1$ and the distribution of axis ratios is substantially broader. At both indices, the distribution remains relatively unchanged as $b_v$ is varied, save for the changes in the number of filamentary objects with $q,~s\lesssim0.1$. These filamentary objects are elongated subhaloes in the process of being tidally disrupted with loosely bound tidal tails. These filaments comprise $\lesssim1\%$ of the subhalo population for $b_v\leq0.05$ at $n=-1$ and for $b_v\leq0.025$ at $n=-2.5$, and increase to $20\%$ for $b_v=0.10$ at both indices. This filamentary population decreases by roughly a factor of $\approx10$ and $\approx5$ as $b_v$ is halved at $n=-1$ and $n=-2.5$ respectively, though the trend is not as clear cut at $n=-2.5$. The main reason for this dependence on $b_v$ is that tidal tails are more likely to be linked with the parent subhalo when using large $b_v$. The decrease in the number of subhaloes as $b_v$ is increased at $n=-2.5$ indicates that, as $n\rightarrow-3$, local phase-space peaks become less well separated in phase-space and more likely to be embedded in a triaxial overdensity. The increasingly triaxial or filamentary nature makes assigning a mass to a phase-space peak non-trivial. Estimates of a tidal mass assuming spherical symmetry will become increasingly incorrect as the distribution of mass around a phase-space peak becomes increasingly triaxial. These results are insensitive to the minimum subhalo particle number cut applied. Again, the differences in morphology may be indicative of a break in the ``universality'' of the subhalo mass function.

\subsection{Mass function}
\label{sec:subhalomass}
\begin{figure*}
    \centering
    \includegraphics[width=0.8\textwidth]{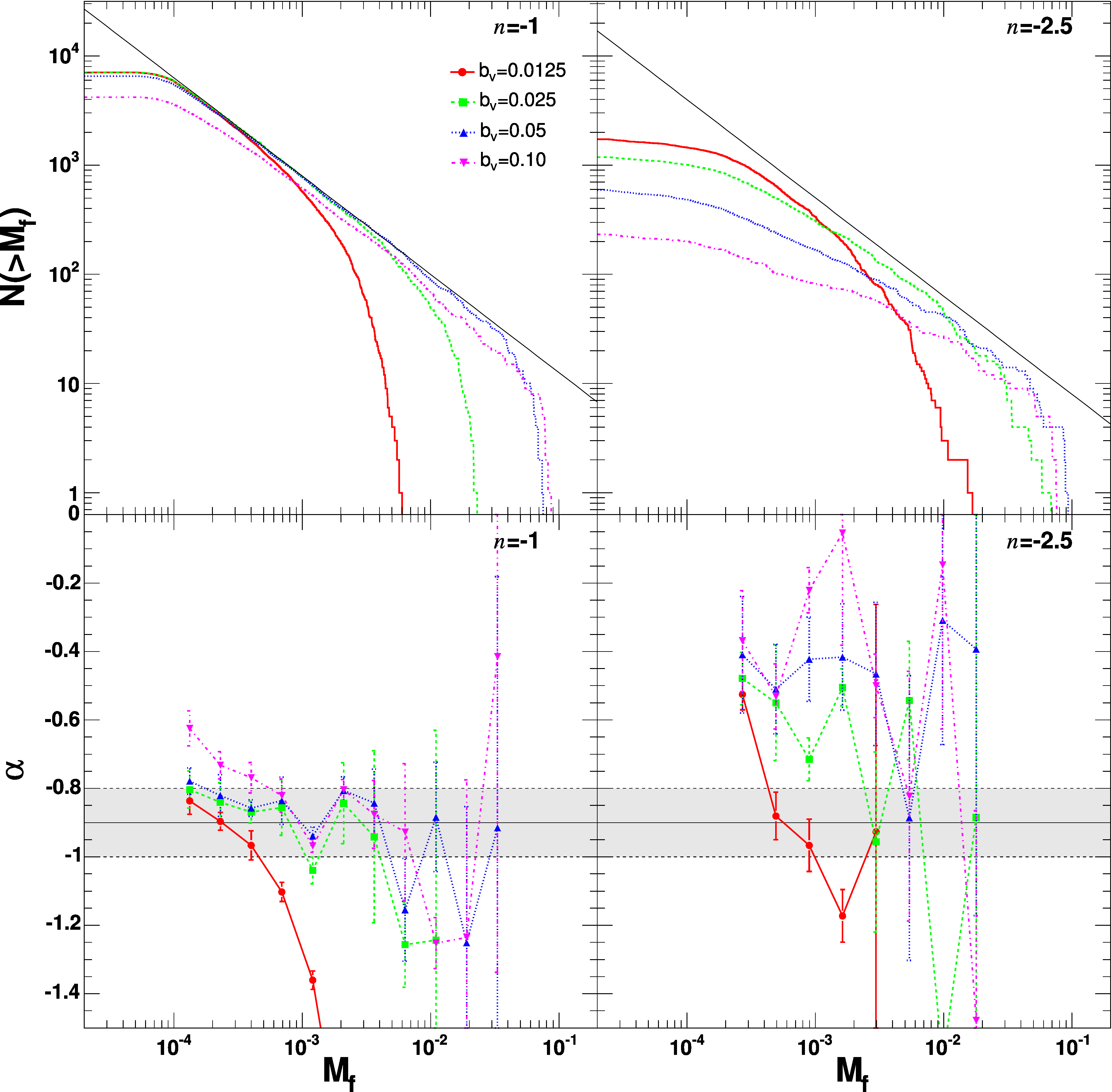}
    \caption{The cumulative subhalo mass function $N(>M_f)$ (top) and logarithmic slope $\alpha$ (bottom) from all haloes used in our analysis. The line styles and marker scheme are the same as in \Figref{fig:submorph}. Also shown in the top panel by the solid thin black line is $N(>M)\propto M^{-0.9}$. The solid grey region outlined by black dashed lines in the bottom panel indicates the various slopes found in other studies with the solid horizontal line denoting $\alpha=-0.9$.}
    \label{fig:submass}
\end{figure*}

We define the dimensionless subhalo mass ratio $M_f\equiv M_{\text{subhalo}}/M_{\text{halo}}$ and show in \Figref{fig:submass} the cumulative subhalo mass function $N(>M_f)$ summed over all haloes in our analysis. We also show the logarithmic slope $\alpha(M_f)\equiv d\ln (dN/d\ln M_f)/d\ln M_f$. Since $dN/d\ln M_f$ is quite noisy, we calculate several logarithmic slopes at a given $M_f$ by varying the spacing used in the five point central difference estimate of the slope. Shown in \Figref{fig:submass} is the average slope along with the standard deviation. In both simulations, $N(>M_f)$ is linear in the log-log plot provided we exclude low and high mass regions dominated by numerical effects. The flattening in the low mass region corresponds to subhaloes composed of fewer than 100 particles and is due to numerical softening effects. The high mass scale, where the distributions begins to steepen, scales roughly as $b_v^3$. This region results from excluding the outer high velocity volumes of a candidate subhalo and the scaling can be understood as follows: $v^2\propto M/R$, $R\propto M^{1/3}$, thus $v^3\propto M$. Hence, more massive the subhalo the larger $b_v$ must be to group particles out to the subhalo's tidal radius. Since we pass subhalo candidates through an unbinding routine, which does not take into account the background, the loosely unbound central regions of large subhaloes will be removed.

\par
In the $n=-1$ simulation, the number of subhaloes is very weakly dependent on $b_v$. The 6DFOF algorithm using $b_v=0.10$ cannot separate subhaloes in the central region from the core and underestimates the number of subhaloes. The small changes in the number found with $b_v\leq0.05$ are due to the fact that the 6DFOF algorithm using these velocity linking lengths tends to only link the central regions of subhaloes with mass ratios of $M_f\gtrsim0.01$, underestimating their mass. In some cases, the algorithm only groups the loosely unbound centres of these subhaloes and they are subsequently removed by our unbinding routine. However, this does not greatly affect the number since there are few subhaloes above the high mass limit imposed by $b_v$. The fraction of mass bound in subhaloes with $10^{-4}\leq M_f \leq 5\times10^{-3}$ is weakly dependent on $b_v$, decreasing from $0.127$ to $0.085$ when going from $b_v=0.0125$ to $b_v=0.10$. This fraction varies from halo to halo by an average of $\approx10\%$ for $b_v\leq0.05$ and increases to $\approx20\%$ for $b_v=0.10$ with no strong dependence on peak height of the halo. The slope of the subhalo mass function is also unaffected by changes in $b_v$ and, neglecting the regions dominated by numerical effects, is consistent with previous results, as indicated by the grey region in \Figref{fig:submass}. Even between the catalogues found using $b_v=0.10$ and $b_v\leq0.05$, where half as many subhaloes are found and the fraction of filaments goes from $0.2$ to $\lesssim0.01$, the slope is only shallower for mass ratios of $M_f\lesssim 2\times10^{-4}$.

\par
The subhalo distribution at $n=-2.5$ is dependent on $b_v$. The number of subhaloes increases as $b_v$ decreases and appears to converge, though we note that few if any bound subhaloes are found using $b_v<0.0125$. The increase in the number found is due to bound filamentary objects found at a given $b_v$ being broken into several smaller triaxial subhaloes at smaller $b_v$, as shown in \Figref{fig:subcand} \& \ref{fig:submorph}. For example, comparing the subhaloes found using $b_v=0.025$ to those found using $b_v=0.0125$, the number of filaments decreases from $1.7\%$ to $0.3\%$ of the population but the number of subhaloes increases by a factor of $1.6$. The apparent convergence is due to there being a finite number of phase-space peaks with local physical overdensities above $\gtrsim1000$. The fraction of mass in subhaloes between $10^{-4}\leq M_f \leq 5\times10^{-3}$ shows a strong dependence on $b_v$, decreasing from $0.035$ to $0.0052$ when going from $b_v=0.0125$ to $b_v=0.10$. In general, the fraction of mass bound in subhaloes is a factor of $\sim3-20$ smaller than at $n=-1$. The relative halo to halo variation is also greater, with the fraction varying by $26\%$ and $96\%$ using $b_v=0.0125$ and $b_v=0.10$ respectively. The slope increases as $b_v$ increases and exhibits fluctuations as a function of $M_f$ that are greater than those at $n=-1$.  Generally, $\alpha\gtrsim-0.9$ within the region free of numerical effects. Given the changes in the number of subhaloes found, the changes in $\alpha$ could be dismissed as being purely numerical. However, similar changes in the subhalo population are observed at $n=-1$ going from $b_v=0.10$ to $b_v=0.05$ yet $\alpha$ is unaffected, whereas at $n=-2.5$ $\alpha$ increases by $\approx20\%$ when going from $b_v=0.05$ to $b_v=0.025$ or $b_v=0.025$ to $b_v=0.0125$. These changes in $\alpha$, while being partly numerical, are nonetheless indicative of the underlying physical issue of defining the boundary of a subhalo.

\par
\begin{figure}
    \centering
    \includegraphics[width=0.44\textwidth]{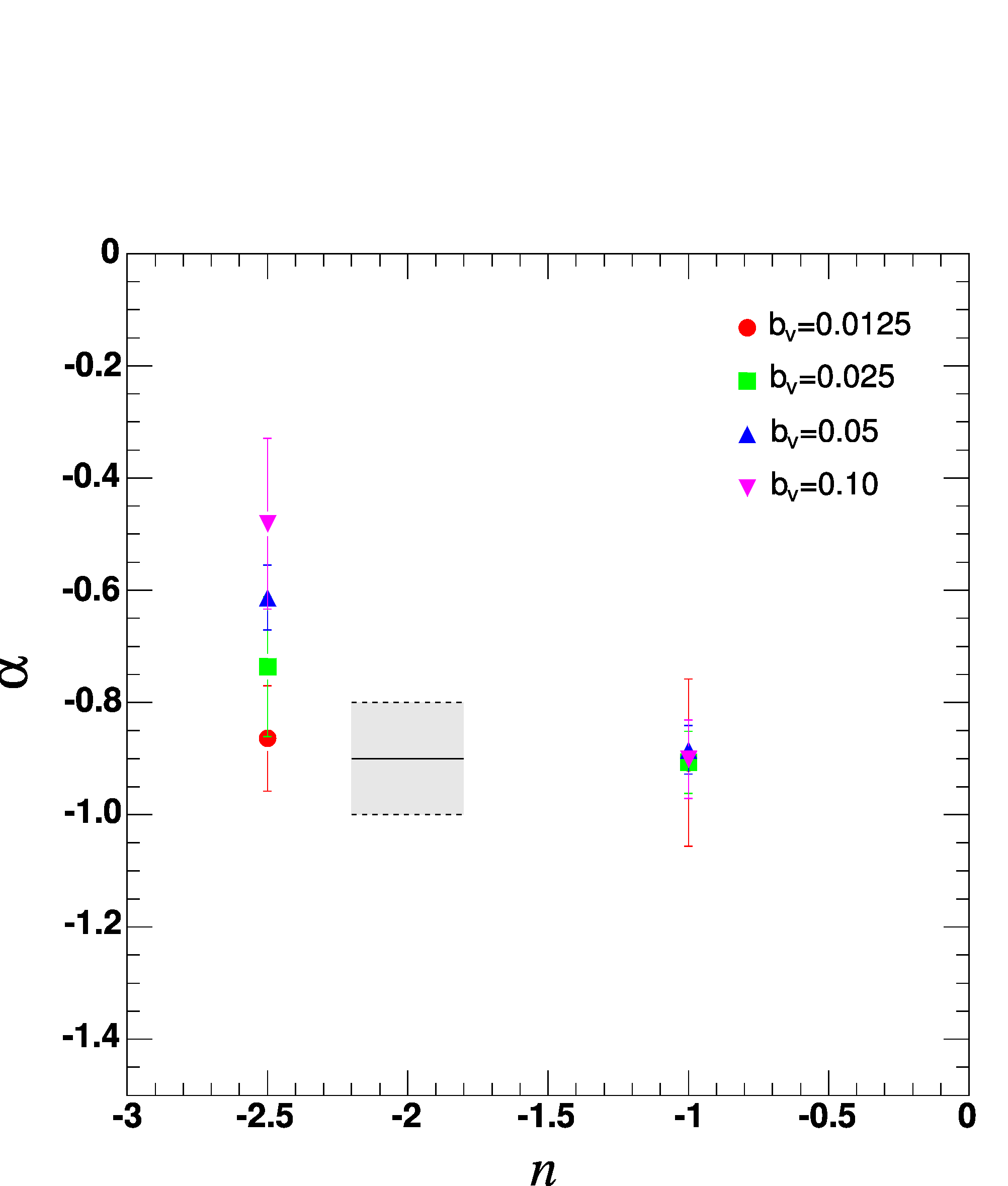}
    \caption{Spectral dependence of $\alpha$. Filled stars, circles, squares and triangles indicate slopes from subhaloes found using $b_v=0.0125,~0.025,~0.05$~and~$0.10$ respectively. Marker scheme is the same as in \Figref{fig:submorph}. The grey region and black lines indicate the same slopes as in \Figref{fig:submass} at $\neff=-1.8$~to~$-2.2$, which correspond to cluster scales down to galaxy scales in the CDM hierarchy. Error bars correspond to the average variation in the logarithmic slope seen in bottom panel of \Figref{fig:submass} within the region fitted by the power-law.}
    \label{fig:submassnspec}
\end{figure}

To compare with previous work, we fit a power-law,
\begin{equation}
N(>M_f)=AM_f^{\alpha},
\end{equation}
to the cumulative subhalo mass function, neglecting the low and high mass regions which are dominated by numerical effects. The mass scale above which subhaloes are artificially truncation due to $b_v$ is very clear at $n=-1$ but not as clear at $n=-2.5$, so we use the $n=-1$ simulation for guidance in defining the range in $M_f$ fitted. We plot the mean $\alpha$ as a function of the spectral index for different $b_v$ in \Figref{fig:submassnspec}. At $n=-1$, $\alpha\approx-0.9$ independent of the choice of velocity parameter in agreement with previous results from $\Lambda$CDM simulations. Examining the four largest haloes in the $n=-1$ simulation, we find the halo to halo variation in the fitted index is $0.07$. Unlike at $n=-1$, the index at $n=-2.5$ exhibits a strong dependence on $b_v$, though it has similar a halo to halo variation of $0.06$. We can reproduce $\alpha\approx-0.9$ by choosing $b_v\approx0.0125$, though in general $\alpha\gtrsim-0.9$. Increasing $b_v$ from $0.0125$ to $0.025$, that is applying higher velocity cutoffs and then imposing a tidal limit, does flatten the slope by $22\%$. These results indicate that the slope of the subhalo mass function depends on $n$, becoming shallower but more sensitive to systematics as $n\rightarrow-3$.

\par
To check for any redshift dependence, we examine the evolution of a subset of our haloes over $0.8$ and $0.5$ $e$-foldings in expansion factors for $n=-1$ and $n=-2.5$ respectively. We also compare haloes between the two simulations at similar redshifts. However, due to the redshift and mass dependence of $\nu$, we cannot at the same time continue to compare similar mass haloes originating from similar $\sigma$ peaks. We find that subhaloes do not have a strong systematic redshift dependence in either cosmology, thus the differences between the simulations remain unchanged. At $n=-1$, $\alpha\approx-0.9$ regardless of redshift and at $n=-2.5$ the value of $\alpha$ and its dependence on $b_v$ remains unchanged. It is interesting to note that $\alpha$ does vary randomly with redshift by $\approx0.05$ and $\approx0.06$ for $n=-1$ and $n=-2.5$ respectively. This variation is of similar size to the halo to halo variation seen, which for the subset studied spans a peak height range of $1\lesssim\nu\lesssim4.5$ and $2.2\lesssim\nu\lesssim3.3$ for $n=-1$ and $n=-2.5$ respectively.

\section{Discussion and Conclusion}
\label{sec:discussion}
One of the most intriguing results from the \citet{moore1999} study of CDM structure is similarity of substructure in galactic and cluster haloes. This result suggests that the subhalo mass distribution is independent of parent halo mass, and possibly independent of shape of the primordial power spectrum. At first glance, this similarity may seem suprising since galaxy and cluster haloes differ by two orders of magnitude in mass. However, galaxies and clusters probe similar effective spectral indices, ($\neff\approx-2.2$ as compared to $\neff\approx-1.8$), and it is the effective index that governs the structure formation process via the relative formation times and merger rates of progenitors.

\par
To explore this issue, we have analyzed substructure in Einstein-de Sitter cosmologies at two widely separated indices, $n=-1$ and $-2.5$, which bracket the indices most often examined in $\Lambda$CDM simulations. In the spirit of \citet{smith2003} we have outlined a set of criteria to minimize the numerical effects which plague simulations of $n<-2$ cosmologies. We chose $n=-2.5$ as it is the most negative index we can simulate that satisfies our criteria, albeit just barely. Another reason for choosing $n=-2.5$ is that it represents a transitional index between hierarchical formation and the singular, perhaps pathological, case of $n=-3$ where structures collapse simultaneously. We evolved our simulations until there were haloes large enough to analyze for substructure, while at the same time not being too significantly affected by finite volume effects. Substructure was analyzed with a number of tools: we examined the phase-space structure of haloes using EnBiD, searched for candidate subhaloes using a 6DFOF algorithm, and examined the morphology and mass distribution of bound subhaloes. All of these methods showed that substructure in an $n=-2.5$ cosmology is different from substructure in an $n=-1$ cosmology. For example, substructure leaves a feature in the volume distribution function whose strength depends on the number of subhaloes and the density contrast of substructure (\citealp{arad2004}; \citealp{binney2005}; and \citealp{sharma2006}). This feature was present in our $n=-1$ simulation but was greatly reduced in our $n=-2.5$ simulation.

\par
The differences in the morphology of bound subhaloes between the two cosmologies are striking and appear to be independent of redshift. At $n=-1$, subhaloes were typically well-defined, spherical overdensities that would lend themselves to the type of analysis outlined by \citet{diemand2007}, where the tidal mass is determined by fitting spherically averaged density profiles to phase-space peaks. At $n=-2.5$, subhaloes were more triaxial and phase-space peaks were more irregular, making it difficult to define a boundary. The triaxial or filamentary nature of subhaloes, along with the large fraction of unbound candidates, indicates that substructure in an $n=-2.5$ cosmology is more susceptible to tidal disruption. These observations suggest that the methods that assume a spherical profile in order to estimate subhalo circular velocities and masses, such as those outlined in \citet{diemand2006,diemand2007}, will become increasingly inaccurate as $n\rightarrow-3$ and might give misleading results.

\par
The logarithmic slope of the subhalo mass distribution also showed a spectral dependence. At $n=-1$, we found $\alpha\approx-0.9$ independent of the the velocity linking length $b_v$ and is consistent with slopes from previous studies. At $n=-2.5$, we found a range of values $-0.9<\alpha<-0.5$ depending on our choice of $b_v$ with $\alpha$ tending towards less negative values with larger values of $b_v$. The sensitivity of $\alpha$ to $b_v$ at $n=-2.5$ is due to the underlying uncertainty in determining the boundary of a subhalo. Our preferred value is $\alpha=-0.76\pm0.07$, found using $b_v=0.025$, for a number of reasons: subhaloes were found over several decades in $M_f$ below the numerical mass limit imposed by $b_v$ and above the mass scales dominated by numerical softening effects; highly filamentary subhaloes made up $\approx1\%$ of the population; and the number of subhaloes appeared to be converging. We also find that the logarithmic slope of the mass function shows no dependence on redshift, though $\alpha$ does vary by $\approx5-10\%$ in time for a given halo. This random variation in $\alpha$ with redshift is of similar size halo to halo variation, which shows little dependence on the peak height of the halo.

\par
Our preferred slope at $n=-2.5$ appears to agree with that of the subsubhalo mass function in $\sim10^9\Msun$ subhaloes studied by \citet{springel2008}, where $\neff\approx-2.4$. However, we note that the subsubhalo mass functions in \citet{springel2008} are quite noisy with a great deal of scatter in the observed logarithmic slope, so this agreement should be treated with caution. This value is not  entirely consistent with that found in \citet{gao2005} and \citet{diemand2006}, who both found results consistent with previous studies of much larger scales. Using SUBFIND, \citet{gao2005} examined substructure in a single high redshift subgalactic halo at $\neff\lesssim-2.5$ originating from an extremely rare $\approx5\sigma$ peak, whereas we examined the logarithmic slope averaged over a number of haloes, albeit they examined their halo over a larger redshift range than we did. At their earliest epoch, this rare peak corresponded to a $2\times10^5\Msun$ halo composed of $\approx2\times10^5$ particles with $\neff\approx-2.7$. Though their halo sampled a more negative index then our simulation, it was an extremely rare peak that only had $\approx35$ subhaloes over a single decade in $M_f$ from $10^{-4}\lesssim M_f\lesssim10^{-3}$. We also note that the logarithmic slope of this rare peak appears to vary by $\sim0.1$ with redshift, though there is no systematic dependence on redshift, similar to the variation we observed. \citet{diemand2006} also examined a single rare halo, though at an effective index slightly closer to $-3$ and at a much higher resolution than \citet{gao2005} study. They noted that the subhalo mass function was sensitive to the parameters used to find candidates, unlike the mass function of larger haloes where $\neff\gtrsim-2$, in agreement with our observations, although they used SKID and we used a 6DFOF.

\par
Our results, when combined with those of previous studies (e.g.~\citealp{moore1999,gao2004,reed2005,diemand2007}; and \citealp{springel2008}), imply that the logarithmic slope, though relativity constant for a given host, is dependent on the host's mass and its associated spectral index. This slope is constant with a value of $\approx-0.9$ so long as $n\gtrsim-2$ but tends to larger values and possibly with increasing scatter for $n\lesssim-2$. This transition in behaviour might be due to the structure formation process. Provided that this process is ``sufficiently'' hierarchical, that is haloes virialize before being accreted or merging, the subhalo mass function is independent of $n$. As $n\rightarrow-3$, haloes do not fully virialize before being accreted and consequently the subhalo distribution is influenced by the properties of the power spectrum. This dependence manifests itself in the boundary of a subhalo becoming increasingly ill defined as $n\rightarrow-3$; phase-space peaks become more indistinct and irregular, resulting in greater variation in the subhalo mass function while also tending to flatten it. Thus, extrapolating the subhalo mass function at galactic scales to predict the number of subhaloes at the bottom of the CDM hierarchy is questionable. It is likely that at $n=-3$ subhaloes become generally impossible to distinguish and, in essence, substructure becomes negligible.

\par
The distribution and internal properties of subhaloes have important ramifications for dark matter detectors. Because the $\gamma$-ray flux varies as the local dark matter density squared, indirect detectors, such as GLAST, will see a stronger signal if there is a great deal of dense substructure. Numerous groups have calculated the boost in flux due to substructure (e.g.~\citealp{strigari2007}; \citealp{pieri2008}; and \citealp{kuhlen2008}) but there are two issues with these estimates. The primary issue is the scale dependence of the subhalo mass function. Estimates assume the subhalo mass function observed in $10^{12}\Msun$ haloes at subhalo masses of $\gtrsim10^{5}\Msun$ applies to the subsubhalo population and extends to the bottom of the CDM hierarchy. However, it is likely that the subsubhalo mass function of $\sim10^{8}~\Msun$ subhaloes will have a form similar to that observed at $n=-2.5$ for $b_v=0.025$ and that the distribution will continue to flatten as one proceeds down the CDM hierarchy. The secondary issue is the use of field halo mass-concentration relations to convert subhalo masses to densities. At small scales, their use is questionable since subhaloes become increasingly triaxial and less distinct as $n\rightarrow-3$. However, this could be  accounted for in the concentration-mass relation such as the one proposed by \citet{eke2001} which attempts to account for the possible effects of $\neff$. Ultimately, as a consequence of these issues, previous results should generally be considered optimistic upper limits. In light of these results we do not attempt to predict the $\gamma$-ray background here and will address this critical issue in a paper in preparation.


\section*{Acknowledgements}

The authors thank the anonymous referee for useful comments on the manuscript. PJE acknowledges financial support from the Natural Science and Engineering Research Council of Canada (NSERC). RJT and LMW acknowledge funding by respective Discovery Grants from NSERC. RJT is also supported by grants from the Canada Foundation for Innovation and the Canada Research Chairs Program. Simulations and analysis were performed on the computing facilities at the High Performance Computing Virtual Laboratory at Queen's University, SHARCNET, Arizona State University Fulton High Performance Computing Initiative and the {\em Computational Astrophysics Laboratory} at Saint Mary's University. 


\bibliographystyle{mn2e}
\bibliography{subhalopaper.bbl}

\label{lastpage}

\end{document}